\documentclass[article]{elsarticle}

\usepackage{algorithm}
\usepackage{graphicx}
\usepackage{epsfig,amssymb,amsmath,bezier}
\usepackage[caption=false,font=footnotesize]{subfig}
\usepackage{verbatim}
\usepackage{url}
\usepackage{algpseudocode}
\usepackage{natbib}

\journal{Computer Networks}

\begin{document}

\begin{frontmatter}

\title{Efficient Cache Availability Management in Information-Centric Networks}
\author[cse-aalto]{Sumanta Saha\corref{cor1}}
\ead{sumanta.saha@aalto.fi}

\author[cse-aalto]{Andrey Lukyanenko}
\ead{andrey.lukyanenko@aalto.fi}

\author[cse-aalto]{Antti Yl\"a-J\"a\"aski}
\ead{antti.yla-jaaski@aalto.fi}

\cortext[cor1]{Corresponding author}

\address[cse-aalto]{Computer Science and Engineering, Aalto University, 02150 Espoo, Finland}

\begin{abstract}
In-network caching is one of the fundamental operations of Information-centric networks (ICN). The default caching strategy taken by most of the current ICN proposals is caching along--default--path, which makes popular objects to be cached redundantly across the network, resulting in a low utilization of available cache space. On the other hand, efficient use of network-wide cache space requires possible cooperation among caching routers without the use of excessive signaling burden. While most of the cache optimization efforts strive to improve the latency and the overall traffic efficiency, we have taken a different path in this work and improved the storage efficiency of the cache space so that it is utilized to its most.

In this work we discuss the ICN caching problem, and propose a novel distributed architecture to efficiently use the network-wide cache storage space based on distributed caching. The proposal achieves cache retention efficiency by means of controlled traffic redirection and selective caching. We utilize the ICN mechanisms and routing protocol messages for decision making, thus reducing the overall signaling need. Our proposal achieves almost 9-fold increase in cache storage efficiency, and around 20\% increase in server load reduction when compared to the classic caching methods used in contemporary ICN proposals.
\end{abstract}

\begin{keyword}
ICN \sep routing \sep distributed algorithm \sep cache
\end{keyword}

\end{frontmatter}

\section{Introduction}

The transformation of the Internet from a medium of communication between two fixed points---a known server and a client---to a more content oriented paradigm warrants a radical change in the architecture. The Internet today can be represented as an eyeball--transit--content model, where data is concentrated on one side and customers on another~\cite{eyeball}. The primary purpose of such a model is data availability irrespective of the source of the data. Information Centric Networking (ICN) is one of the recent future Internet research focuses~\cite{dona,ccn} which takes into account this change and solves this architectural problem from within the network instead of using application overlays.

While ICN is a rather clear and well understood concept, the exact realization of many core parts, e.g. caching, is yet to be well understood~\cite{icn2011}. While many proposals agree that caching is an integral part of ICN~\cite{4ward, psirp, sail} and propose several solutions to use in-network caches~\cite{diallo2013, kurose2009}, none of them focuses on the storage efficiency of the cache, that is how the data is distributed across the total cache space available in the network. Most of the caching solutions around concentrate on the efficiency of the cache in terms of content delivery latency. However, another aspect of caching is not recognized by many, that is, the storage efficiency aspect where the total cache space in the network is used efficiently to increase the availability of the total amount of data. Considering that ICN poses no limitation on where the data can be served from, better cache storage efficiency can result in a much better data retention capability for the network. The network can retain the data objects (please note that, we do not mean application layer objects here, rather it can be any unit of data such as packet chunks as used in, for example, CCNx~\cite{ccn}) in the in-network-cache\footnote{In-network-cache refers to a caching scheme where caching is integrated in core network elements, such as routers, instead of being overlayed as HTTP caches} based on the interest level, and thus keep it available for the clients even in the absence of the server. For this to happen, the total cache space in the network has to be utilized efficiently so as to retain as much data as possible. However, current solutions (e.g. CCNx~\cite{ccn}, PSIRP~\cite{psirp}) only cache along the data delivery path with suitable cache replacement policies (e.g., the Least Recently Used (LRU)~\cite{cache_policy}), or cache in predefined off-path locations only for popular objects~\cite{psirp3}. While LRU is very good at maintaining a healthy state of local caches, it is a non-cooperative solution and thus cannot make decisions which would help the network as a whole. On the other hand, cache cooperation strategies, where distributed caches talk to each other to share the cache catalog and to synchronize the caching responsibility, experienced major problems in earlier tries and were considered unfeasible~\cite{cache_coop}. Thus, such solution is missing in the current proposals that can make decisions to help the caching scenario of the network as a whole, while keeping the complexity feasible.

As discussed in~\cite{hotnets2011}, imagine that we have a route consisting of $20$ hops, and each router on the way is able to store up to $10^8$ objects (or data chunks) while the total number of objects is $10^{12}$. Hence, we may store only $2\cdot 10^{9}$ objects in the most idealistic case and the remaining will be requested from the source. Caching $0.001\%$ or less of the total content would work if a minority of contents would be highly popular. However, as shown in earlier works~\cite{kazaa, gill2007youtube} and recently verified on CDN workloads~\cite{fayazbakhsh2013less}, the traffic profile in current Internet follows a heavy-tailed or Zipf distribution (and often Zipf-Mandelbrot distribution) in request popularities. In this kind of traffic distribution, a large portion of traffic remains in the tail of the distribution which never ends up retained in the cache. Requesting a majority of the data in the heavy tail of the distribution from the origin makes the efficiency of caching and their use in ICN rather questionable as it downgrades the ICN to the current Internet working principle (Web caches, Content Delivery Networks (CDN), etc). Moreover, recent works such as~\cite{fayazbakhsh2013less} showed that optimizing caching techniques to gain on parameters such as latency is not realistic and results in almost similar performance as very simple caching techniques (e.g. edge caching). In addition, the same work demonstrates that the latency improvement attributed to universal caching compared to simple edge caching is only $25\%$ considering that edge caching actually uses only half of the total cache capacity to reach this performance. For these reasons, while most of the cache optimization works have concentrated on the cache efficiency parameters such as latency or network traffic, our focus in this work is on the cache availability parameter. Instead of trying to gain on latency, we have focused on keeping as much data as possible in the cache and thus achieve a high rate of cache hits. Although this approach can penalize the network with worse latency resulting in more network traffic, the measurements show that the deterioration can be quite acceptable. Moreover, it is possible to improve the latency problem by restricting the detour process with some finite bound and by fine-tuning the retrieval algorithm to work within the bounds while keeping the cache hit ratio on a high level. However, we have not yet explored the bounded version in this work. 

In this paper we present a novel ICN routing protocol which ensures better network-wide cache utilization, allows data servers to delegate part of their load to caching routers, and increases data availability in the network. The novel part of the proposed design is how a lightweight cooperation is integrated with the ICN architectures. While cache cooperation techniques mostly failed due to excessive and prohibitive signaling costs, the proposed design achieves such cooperation among distributed caches with very low signaling overhead. Additionally, the design is gradually deployable, which makes the deployability challenge---which any new design faces---much easier. 

In this work, we propose a distributed architecture that can be deployed gradually. We propose first an AS-local solution, where the modifications are bound within a single AS. The routing algorithm assigns a subset of the universal data space to each of the caching routers and routes data requests according to the assignment. In the next iteration, this idea is extended to the whole Internet by introducing a parameter called ``AS-interest'', and the final iteration extends the idea further to route the packets through a non-default route to hit the interested ASes. We have compared the proposed scenarios with two of the current caching schemes used in ICN networks: CEE or Cache Everything Everywhere, which caches every piece of data in all the nodes (used in most of the recent ICN architectures); and ProbCache~\cite{psarasprobabilistic}, which probabilistically decides on where to cache on the path. The proposed architecture shows ~20\% improvement in server hit ratio, achieves almost 95\% cache storage efficiency (almost 9-fold increase), and demonstrates similar improvements in various other parameters such as cache eviction, etc. compared to CEE or ProbCache. Part of the preliminary results of the proposal are presented in~\cite{myinfocom}. This version of the paper extends the preliminary version with more extensive performance evaluation, and the non-default path scenario. In this version, we looked at the possibility of using non-default AS-path for distributing cache and the relative performance of that. We have measured the performance also with a real world topology extracted from the CAIDA~\cite{skitter} dataset to demonstrate the scalability of the proposed algorithm. 

The rest of the paper is organized as follows. We first introduce related literature in Section~\ref{sec:related}, followed by Section~\ref{sec:problem} which illustrates the problem with an example. Then a more detailed description of our solution is presented in Section~\ref{sec:sol}. In Section~\ref{sec:perfeval}, we present a performance evaluation to back our architecture up, followed by a discussion about the architectural design in Sections~\ref{sec:disc} and \ref{sec:concl}.

\section{Background and Related Work}
\label{sec:related}
ICN or Information Centric Network has emerged as a new paradigm for the underlying architecture of the Internet. In recent years, a wide variety of ICN architectures have been proposed with different feature sets~\cite{ccn, netinf, psirp, triad, dona}, and a common principle: the network primitives such as routing, forwarding, etc should be based on the content name instead of the server address.

The idea of ICN can be attributed to a very early work named TRIAD~\cite{triad}, which introduced publish-subscribe as a basic Internet paradigm (inspired from~\cite{oldpubsub}). A long time after TRIAD was proposed, networking researchers started to realize the validity of the work, which resulted in a widespread interest on similar topics, producing different ICN architectures such as CCN~\cite{ccn}, 4WARD~\cite{4ward, 4ward2, 4ward3}, PSIRP~\cite{psirp, psirp2, psirp3, psirp4}, SAIL~\cite{sail, sail2}, and COMET~\cite{comet, comet2, comet3}. One common aspect of almost all of these proposals is universal caching with minor variants. That is, instead of having few specialized caching points across the network as an overlay, caching is implemented by all the ICN nodes. 

CCN~\cite{ccn} uses interest messages and self-identifying names to route data request. It uses a cache everything everywhere (CEE) approach, and relies on the cache eviction methods such as Least Recently Used (LRU). NetInf~\cite{netinf}, an architecture developed as a result of multiple projects such as 4WARD~\cite{4ward, 4ward2, 4ward3} and SAIL~\cite{sail}, allows several caching mechanism such as cooperative caching with the help of a centralized system, or a more proactive caching where caches subscribe to contents themselves. In PSIRP~\cite{psirp, psirp2, psirp4} caching is limited to the transmission path, and registered within the name resolution system. PURSUIT~\cite{psirp3}, which is a continuation of PSIRP does not cache on path to allow asymmetric routing. However, still the caching is concentrated on predefined areas. None of these architectures propose any method for the efficient use of the available cache space, resulting in high dependency on the server for data delivery.

Apart from complete ICN architectures, researchers have also proposed caching solutions for generic ICN networks. Cooperative caching solutions provide the most optimal results, however, the signaling cost for this kind of solutions is prohibitive and impractical for live networks~\cite{cache_coop, fayazbakhsh2013less}. Recently, the aggressive cache everything everywhere strategy of CCN~\cite{ccn} was analyzed and questioned in~\cite{hotnets2011}, and a few works have considered more selective caching techniques for ICN~\cite{psarasprobabilistic, chai2012cacheless}, which probabilistically determine whether to cache on path or not to gain better cache and server hit ratio by utilizing the concept of betweenness centrality. They still deal with on-path caches, and cannot take advantage of off-path (not in the default path from client to server) contents. On the other hand, authors in~\cite{wang2011could} analyzed the caching problem in the light of linear programming model and proposed a novel caching policy which takes into account the benefit of cache hit in the total cache governance area. Additionally, in~\cite{diallo2013} the authors have proposed a new publish/subscribe framework for better scaling. Besides presenting the pub-sub paradigm and its modifications, the work also discusses about the effect of caches on the pub-sub mechanism. However, this work also discuss only on-path caching and thus cannot utilize the power of off-path caching. 

Numerous research proposals addressed cache positioning and cooperation for web caching. Seminal work done by Karger et. al.~\cite{consistenthash} has introduced the idea of consistent hashing for distributed placement of caches, which was followed by proposals (such as Multicache~\cite{multicache}) based on distributed hash tables (DHT), overlay services, and so on. Multicache, for example, builds its caching mechanism on Scribe, which is built on top of Pastry that uses DHT. CoralCDN~\cite{coralcdn} uses DHT for CDN, which is another form of caching. Although we have used a similar approach to DHT in our work to distribute the caching responsibility in a deterministic way, our novelty lies on its application to the ICN architecture and integrating the decision in the routing scheme. Recent work done by~\cite{saino2013hash} and~\cite{sourlas2013cache} touch similar caching problem in ICN, however the scope and the focus of those works are different than ours and they can be thought of as complementary to this work. Authors in~\cite{saino2013hash} proposes a very similar process of distributing the cache within one AS as our previous work~\cite{myinfocom}, however, the focus of the work is to improve the hash function with which such intra-AS cache spread can be achieved, while our work extends the idea to inter-AS cooperation. In~\cite{sourlas2013cache}, the authors present a dynamic programming based approach to modify the routing in ICN to follow the cheapest transportation path based on their popularity, locality, and cache state. 

\section{In-network Caching Problem}
\label{sec:problem}
The ICN design for the future Internet depends on in--network--caching and network wide data availability for distributed data delivery\footnote{With some exceptions such as CCNx\cite{ccn}, which, although having ``content store'' as in-node cache, claims that it does not use that for caching, rather just as packet buffers for queueing and jitter elimination purposes.}. Some of the prominent ICN architectures, such as PSIRP~\cite{psirp}, relies on the network-wide data availability to feed the data request from the clients. The availability of the server is not a critical factor for such a network. As long as there is a healthy amount of interest for the data, it is retained in the network caches, and even in the absence of the server the data is served from the cache. To achieve this, the network cache space needs to be utilized efficiently so as to retain as much data as possible without heavy duplication. However, the needed cooperation for optimal data spread and positioning is not feasible for a network with a scale as large as the Internet. Thus, a radical approach is needed to achieve this target.

\begin{figure}[t]
\centering
\subfloat[Topology]{
    \label{fig:ex_topo}
    \includegraphics[height=100pt]{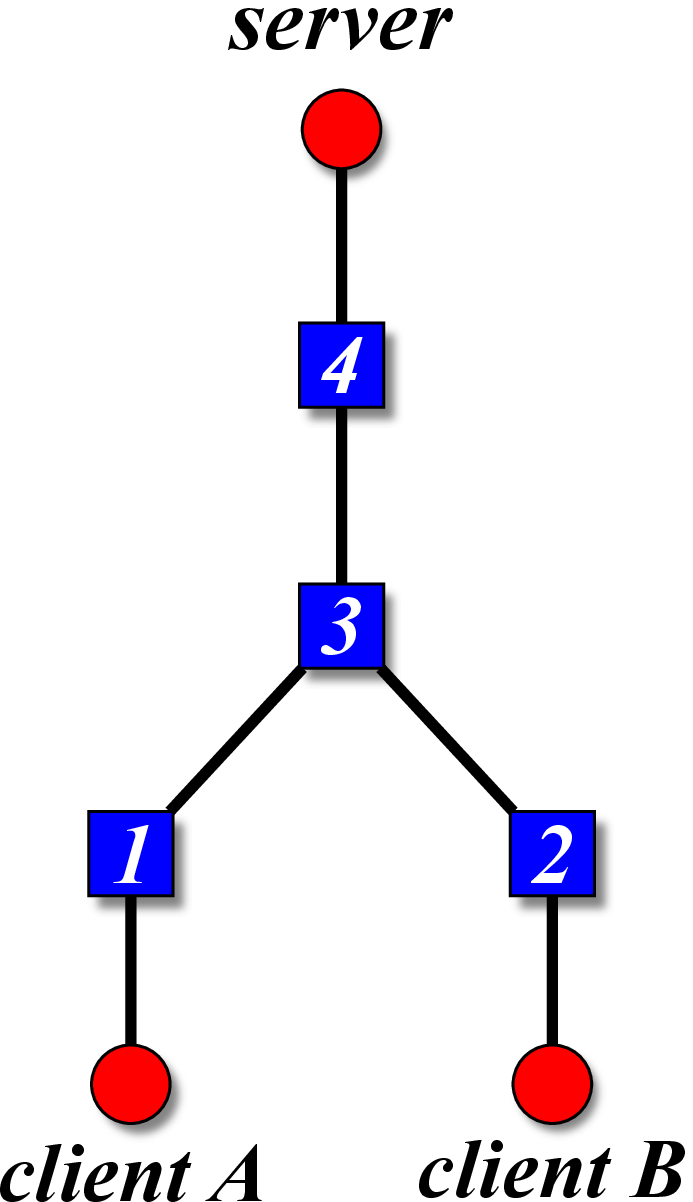}
}
\raisebox{3cm}{\textbf{$\Longrightarrow$}} 
\subfloat[Case 1]{
    \label{fig:ex_case1}
    \includegraphics[height=100pt]{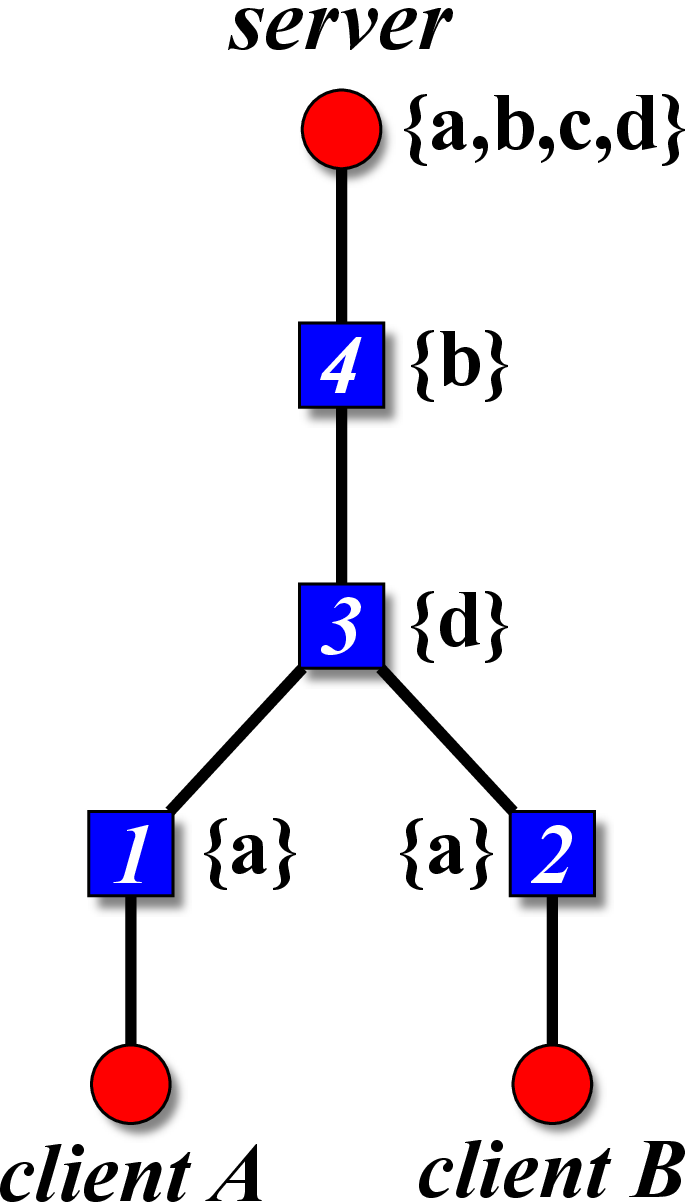}
}
\qquad
\subfloat[Case 2]{
    \label{fig:ex_case2}
    \includegraphics[height=100pt] {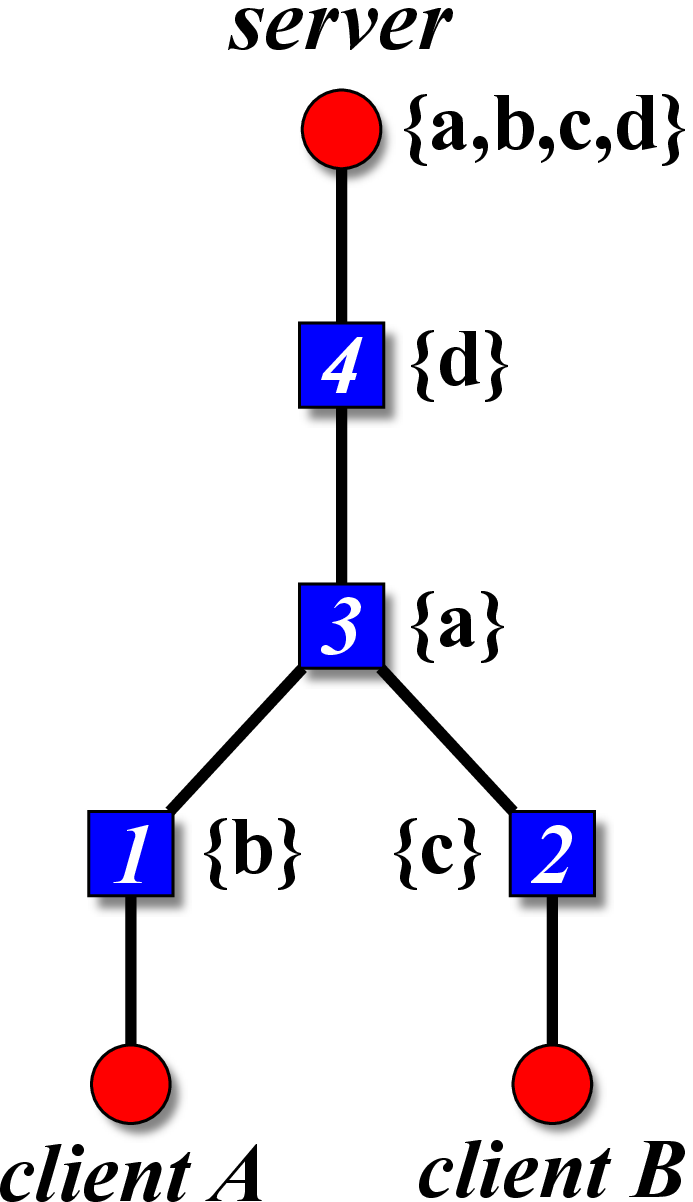}
}
\caption{Caching replacement policies: Case 1. Non Cooperative, Case 2. Cooperative}
\label{fig:example}
\end{figure}

\subsection{Problem}
Our goal in this work is to make efficient use of the cache space available in a network to better serve the content requests, without extensive signaling and excessive server load.

\textbf{An example.} Before focusing on the problem, let us take a look at one elucidatory example. Consider the network set-up shown in Fig.~\ref{fig:ex_topo}. In the network we have 4 objects $\{a,b,c,d\}$ and four content routers (numbered as $1,2,3,4$). Each of which may cache only one object at maximum. Topology has one server (containing all the objects) connected to cache router 4. Let us consider that the objects are sorted by popularity order, i.e., request rates $r_a>r_b>r_c>r_d$ (also for simplicity let $2\cdot r_d\ge max\{r_b,r_c\}$). The client $A$ requests $a,b,d$ and client $B$ requests $a,c,d$.

The structure of caches we present here completely mimics CEE and ProbCache with at least LRU policy. The object named $a, b, c, d$ in practice will be a series of object that are persistent at least once (they are requested while already in a cache, i.e., a hit event happens). Their counterpart are completely transient objects $a’,b’,c’,d’$ also present in the caches, however does not play any role except that they occupy the memory. The difference between them we can define using interarrival time $N_r$. In deterministic case if $N_r(a, t)< \tau$ for some moment $t$ and some threshold $\tau$ then $a$ is persistent after moment $t$, otherwise a transient object. If we define the process stochastically, the condition can be set as an expectation $E\left[N_r(a, t)\right]< \tau$. In this example, for simplicity, we removed the time from the parameters as it is one-step of the time example, we also removed the condition and directly defined persistent elements at this moment of time, i.e., $a,b,c,d$ as well as we make those elements consisting of single object per cache. 

If we consider non-cooperative optimal placement then the routers will cache the objects as shown in Fig.~\ref{fig:ex_case1}. Let the cost for each link usage be $1$, then the total cost will be $2\cdot r_d+2\cdot r_b+3\cdot r_c$. Let us imagine another placement of the objects as depicted in Fig.~\ref{fig:ex_case2}. The total cost for this case will be $2\cdot r_a+4\cdot r_d$. Now, let us find a condition for which the second placement is more optimal than the first one, i.e. $2\cdot r_d+2\cdot r_b+3\cdot r_c>2\cdot r_a+4\cdot r_d$. This condition holds when $r_b+ 1.5\cdot r_c > r_a+r_d$, finally applying condition on $r_d$ ($2\cdot r_d =r_c+2\epsilon$ for $\epsilon=\frac{r_b-r_c}{2}>0$),  we get $r_b+ r_c > r_a+\epsilon$ is the sufficient condition when cache placement in Case 2 is preferable over Case 1. Thus, $r_a$ should be much more popular than $r_b$ and $r_c$ (if $r_b$ is close to $r_c$). This example may be generalized to the non-optimal behavior of non-cooperative network caches, where rarer objects are placed in common paths, while more popular objects go to the edges. Moreover, the popular objects are massively cloned in the network.

One of the main findings of the example is following. It is obvious that there exist trade-off between the cache efficiency in terms of availability and the cache efficiency in terms of hop count or network latency. However, here we have shown that the trade-off is not strict; in some cases, it is possible to increase availability while decreasing the hop count at the same time, what we show with Case 2. On the other hand, the Case 1 example shows how many cache policies try to implement the optimal behavior of the distributed cache system. Thus, increase of availability does not necessarily increase network latency and our task in this work is to find a way to dramatically increase the availability, with controllable impact on the network latency or network traffic efficiency. 

\textbf{Problem statement.} The example brings out the core problem of non-cooperative policies: (a) policies are not aware of the cached objects outside the default path (which is also true for recent semi-cooperative caching approaches such as~\cite{psarasprobabilistic, chai2012cacheless}), (b) policies do not work well for flatter request rates for the objects. 

Given the observations stated previously, our goal is to do a controlled distribution of data objects in order to keep the number of redundantly duplicated objects controllable, while retaining the benefits of replacement policies. The latter is needed to take away the heavy tail of almost unused objects, while the former ensures that the network retains as much data as possible without unnecessarily maintaining multiple copies of the same one. Several studies (e.g.~\cite{hotnets2011}) have argued how the proposed ICN methods are unable to cache the heavy tail of the Internet traffic, and thus do not provide any better benefit than simple edge caches (e.g. CDN). However, with the mechanisms presented in ICN approaches, it should be possible to efficiently use the available cache space of the network to address this problem. To address this problem, we concentrate on data availability to retain as much of the heavy tail as possible within the cache space available. Due to this argument, please note that, rather than placing the data objects (or data chunks) in an optimal position, we focus more on spreading the data objects sufficiently to get the best possible cache hit ratio. This ensures that the network is able to retain the most possible amount of data objects in its cache, and deliver when necessary.

\subsection{Design Decisions and Alternatives}
In this work, our goal is to maintain a controlled distribution of data objects across the network. However, it is obvious that any kind of cooperation between routers themselves, especially on what--is--cached--in--which--router basis, produces much additional signaling traffic and memory states. Moreover, attempts at such cooperation failed previously~\cite{cache_coop}. Therefore, we decided to omit any explicit cooperative algorithms in this work. On the other hand, recall that ICN paradigm has one important property: all the objects have stochastically unique IDs~\cite{ccn}. Hence, instead of placing caches on the data path in a cooperative way, we decide on the caching position for a particular ID, and route based on that. 

For the routing we need to know the network topology beforehand, for example, to extract cooperative routers. Generally, it can be done as a DHT with some extensions; However, we believe that the use of overlays such as DHTs significantly reduces the efficiency of ICN. On the other hand, newer proposals for routing algorithms, such as Pathlet~\cite{pathlet}, allow propagation of such information about the routing topology possible. Please note that, our approach is not restricted to the Pathlet architecture, rather it can be used with any routing protocol (such as BGP extensions) which allows piggybacking of arbitrary message over the already scheduled routing messages. The intention is to disseminate necessary information of our caching algorithm via the routing messages, and not increase the signaling load of the network. For related complexity analysis of such routing messages passing, please refer to algorithms such as Pathlet~\cite{pathlet}. 

The proposal requires that, the routing algorithm used for the network will assume an extra responsibility where the ASes provide their cache interest as supplementary information. This information is not the expensive cache state of the AS, rather the range or subset of the object-space that the AS is interested to cache. For example, let's assume that the data objects are numbered starting from 0 till 1000. In that case, an AS can express its interest in all the objects having indices starting from 50 to 100 using a consice format such as [50, 100]. This interest-range specification also allows interest aggregation (e.g. some AS receiving [100-200] and [200-500] from two other ASes can forward an interest range of [100-500] to the next AS). Routers on the path utilize any basic cache replacement policy (e.g., LRU or LFU). This allows to make cooperation intrinsic in the network, without much excess in signaling traffic. Along with this, operators can resort to intra-AS local cache placement decisions which can help clustering data objects to different regions of the AS, allowing more unique objects to be stored locally, while keeping the decision process cheap.

\begin{figure}[t]
\centering
\subfloat[Routing through a single AS path.] {
    \raisebox{0.8cm}{
        \includegraphics[width=0.40\columnwidth] {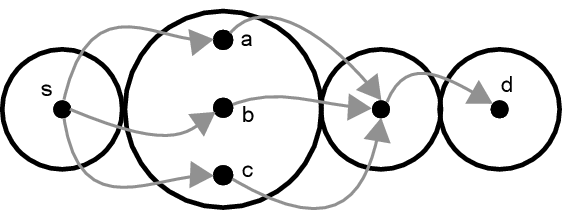}
    }
    \label{fig:as_single}
}
\qquad
\subfloat[Routing through a set of optimal AS paths.] {
    \includegraphics[width=0.40\columnwidth] {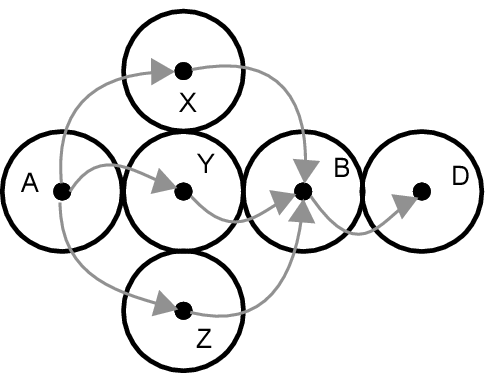}
    \label{fig:as_multi}   
}
\caption{Two types of cache-aware routing in ICN}
\label{fig:as}
\end{figure}

\begin{figure}[t]
\centering
  \includegraphics[width=0.65\columnwidth] {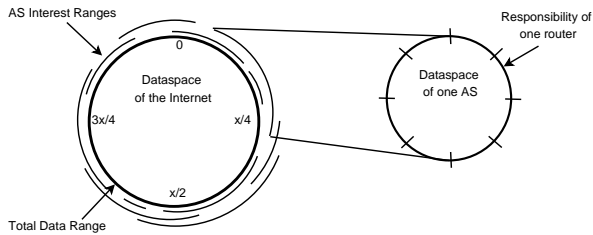}
    \caption{Example data interest of the ASes, and data responsibility of the routers}
    \label{fig:data_range}
\end{figure}

\section{Locally Cooperated Caching Policy}
\label{sec:sol}

The caching scheme works in two hierarchical steps. In the elementary level, the ASes independently modify their internal routing to increase the internal cache efficiency. In the second level, we focus on how the AS--path is selected to find the requested data in the network. We evaluate and compare alternatives in AS path selection and report the result in next sections.

\begin{algorithm}
\footnotesize
  \caption{Proposed caching algorithm}
  \label{alg:algorithm_phases}
  \begin{algorithmic}
    \Statex \Comment {//Each router knows the topology and interest range using algorithms such as pathlet}
      \If{ ($req\_pkt$ not contains $AS\_path$) }
        \State Compute all possible shortest $AS\_path$ from $Requester$ to $Server$
        \If{ ($algorithm$ == SCENE1) }
          \State $AS\_path$ = Select the first shortest path
        \ElsIf{ ($algorithm$ == SCENE2) }
          \State $AS\_path$ = Select the first shortest path containing an interested AS
        \ElsIf{ ($algorithm$ == SCENE3) }
          \State $AS\_path$ = Select a path going through an interested AS (might not be the shortest)
        \EndIf
      \EndIf
      \For{ (each $AS$ in $AS\_path$) }
        \State Calculate $designated\_router$ for $req\_pkt$ (e.g. using consistent hash)
        \State Route to $designated\_router$ from $Border\_router$
        \State Check cache in $designated\_router$ for cache hit
        \If{ (Cache Hit) }
          \State break
        \Else
          \State Route to $Border\_router$ of next AS
        \EndIf
      \EndFor
      \For{ (each $AS$ in reverse $AS\_path$ starting from cache hit) }
        \State Calculate $designated\_router$ for $data\_pkt$ (e.g. using consistent hash)
        \State Route to $designated\_router$ from $Border\_router$
        \If{ ($Data\_pkt$ not in cache) }
          \State Cache the data
        \EndIf
        \State Route to $Border\_router$ of next AS
      \EndFor
  \end{algorithmic}
\end{algorithm}
\normalsize
      
\textbf{Default AS-path routing (SCENE1).} We assume that some, if not all, ASes support data ID-based routing\footnote{In ICN the ID may be thought of as a pair $(P,L)$ as in DONA~\cite{dona}, however, we are not limited to only this type of ID.} on a set of internal routers. Let us assume that there is a default AS path for a server-client pair. In our proposed architecture, every packet which transits through an AS is forwarded through virtually-different internal paths to an outgoing border router (See Fig.~\ref{fig:as_single}). This solution allows to cluster the requests intra-AS on a router basis along the default-AS-path. 

In our implementation, we use a hash function to select a \textit{``designated router''} for each data ID. The hash function maps the global data-ID space to the number of routers inside the current AS [e.g. $hash(DataId) = LocalRouterId | DataId \in GlobalIdSpace$]. The selected router for a particular data ID is then the only router to cache the particular data ID, and the data path for that particular object is always from the incoming-border router to the outgoing-border through the \textit{``designated router''}\footnote{Obviously, we also require for the data to propagate through the same path back; as the hash function is unique within the AS, this condition holds for the whole flow going through AS in a 3-hop manner (i.e., in-border does the ID--to--designated-router hash-based translation; next node (designated-router) do the actual caching and forward the object to the out-border). As an alternative the mechanism may be implemented similar to~\cite{http}.}(simple depiction in Fig.~\ref{fig:data_range}). To avoid link or router failures, hash functions similar to~\cite{consistenthash} can be used. A concise representation of this can be found in Algorithm~\ref{alg:algorithm_phases}.

The use of simple hash fucntions for this purpose has the advantage of being one of the most efficient solution for this purpose as this allows the cache spreading mechanism to be local, distributed, and lightweight. When the operation just requires that a certain caching router takes the responsibility of caching a certain range of data objects without increasing the signaling traffic of the network, a local hashing function provides a very lightweight and fast distributed solution. Recent works, such as~\cite{saino2013hash} and~\cite{myinfocom} have used similar solution to solve this problem.  
 
This first step in clustering the traffic is applicable even in the current Internet. The routers on the path may implement any cache replacement policy.

\textbf{Alternative optimal AS-paths (SCENE2).} Multipath routing schemes such as Pathlet~\cite{pathlet} or similar BGP extensions allow us to freely select from a set of different AS paths. Consider that we start to decide which AS-path to pick among a set of AS paths, where each path is optimal (by AS-count) and the use of one excludes the use of another (Fig.~\ref{fig:as_multi}). To distinguish among optimal paths, we need to obtain some other metric which is in line with our data ID based routing policy. Without loss of generality, assume that each AS together with its own AS-number propagates boundaries of data IDs which it would like to cache (i.e., in form of a continuous sector $[a,b]$; see Fig.~\ref{fig:data_range})\footnote{The complexity of such information dissemination can be found from routing algorithm proposals such as Pathlet~\cite{pathlet}}. Note that, any kind of allowed hash has to have a function-based map to a sector, otherwise (if there are multiplicity of such sectors) the network may be polluted with this information. Thus, any AS should disseminate those sectors with own path updates, the size of a sector is based on the number of internal cache capacity. Whenever AS receives a request with an ID from its own sector $[a,b]$ it should be able to forward the request through internal routers responsible for the ID (i.e., forward to router $i$ if $hash(ID)\in r_i$), and in case of no hit, forward the request farther. 

This scenario facilitates the localization of data one step further than that of default path by taking into account AS interest. It prefers an optimal AS path which contains an interested AS in it (see Algorithm~\ref{alg:algorithm_phases}). Thus, it creates possibilities of spreading around data object by utilizing multi-path, and makes it possible to integrate CDN-like functionality right into the network without using any overlay.

\textbf{Alternative non-optimal AS-paths (SCENE3).} We then explore yet another possibility of cache distribution by allowing data traffic to follow a possibly non-optimal AS path to the data server, and thus increasing the possibility of a cache hit. Similar to the previous method, the ASes disseminate their interest range via routing messages, and the data requester picks the nearest interested AS as an intermediate stop when routing the data request. Thus, the data request travels from the originator to the intermediate interested AS to the data server. If the data is found already in the interested AS, then it is served from there. This also allows to further develop the CDN possibility introduced in previous scenario. With the redirection possibility, ASes can act as a CDN for different sets of data by expressing its caching interest.

\textbf{ICN-aware routing domain.} A overall system architecture view can be extracted from the experiments: The routing domain has a set of interconnected ASes. These ASes propagate updates on the objects they are interested in caching (as compact sector-like representation $[a,b]$) using Pathlet-like~\cite{pathlet} protocol. Based on the update messages, which are piggybacked with the routing messages to avoid extra signaling burden, the protocol generates a map of the nearby ASes and object ranges they are interested in. Any local client requesting for data with some ID, forwards the request to its own edge-router, which then generates possible AS paths towards the guaranteed destination (e.g., as in DONA architecture~\cite{dona}) based on the conditions mentioned in previous sections.  Within each AS on the path, the ingress edge router, using a local caching hash function, forwards the request to a local caching router $r_i$ where $hash(ID)\in r_i$. If data-hit occurs then the reply goes back to the sender through the reverse path, otherwise the data should be forwarded to the border router towards the next AS on path.

The cooperation between the ASes are not mandatory, and only the interested ASes can join the cooperation. Obviously, the more ASes express their interest and cooperate, the better cache spread we are going to get. However, this is not the focus of the study to guarantee the whole content space to be cached. Rather, the work demonstrates that in favorable condition, with adequate amount of ASes cooperating, it is possible to cache much more than the usual caching techniques. The economic feasibility of such caching cooperation of course depends on the policies the ASes have among them. For example, the lower tier ASes (such as tier 2) have the incentive of creating transit links among themselves and use the interest range to avoid going through a higher tier AS. There lies the incremental nature of the proposal. While the method can start from only one AS, where the AS itself tries to distribute the cache contents inside it (SCENE1), it (the method) then continues to evolve without affecting usual Internet operations to inter-AS cooperation. In this phase, the interested ASes join the cooperation and use their existing policies to decide on the caching range. Obviously, with small number of ASes cooperating, the amount of cache distribution will be smaller than a global cooperation, however, it will still increase the cache availability and this will get better as more ASes see the benefit and join the group. 

\section{Performance Evaluation}
\label{sec:perfeval}
We performed a two-level experiment to test the scenarios. On the first level, our focus was to analyze all aspects of the proposal in a controlled environment that provides us with the flexibility to test the proposal in diverse situations and predict its behavior in different scenarios. For this, we used a custom built simulator that maintains the inter-AS and inter-router links among the nodes, and can compute all possible paths between two nodes (for a more detailed description, see~\ref{appn:simulator}). The topology was generated using the BRITE\footnote{http://www.cs.bu.edu/brite/} topology generation tool. The size of the topology was varied under different testing scenarios, and will be mentioned in relevant sections.

In the second level of tests, we used a real-world Internet topology as measured by CAIDA in the Skitter database~\cite{skitter}. The primary focus of these tests was to make sure that the proposal scales and is usable in the Internet. Thus, not all the other protocols were implemented for this level of simulation. In addition to that, these experiments also demonstrates how the performance differs in larger setup for the different phases of the proposal. Total number of ASes in the simulation was around 32000, and for each AS, the number of routers inside the AS was taken from the CAIDA measurement files~\cite{skitter}.

We generated content requests from a fixed pool of content identifiers, and the content popularity was dictated mostly by the Zipf-Mandelbrot (Z-M) distribution \cite{mandelbrot_1967} with parameter $\alpha=0.8$ and $q=5$ to capture the worst case of fairly unpopular data~\cite{rossicaching}, along with the more flattened nature of current data distribution~\cite{hefeeda2008traffic, fayazbakhsh2013less}. 

Although usually the primary target of any caching mechanism is to reduce the hop count and the load on the server, we have concentrated on another aspect of content-centric network along with them: the content spread. It indicates how well unique objects are spread across the network. A well-spread network allows the maximum amount of data to be stored in the network by reducing duplicacy. We have compared the proposed algorithms with the traditional method of content caching (i.e. cache everything everywhere or CEE), and ProbCache, a probabilistic caching algorithm for ICN~\cite{psarasprobabilistic}. Please note that, the results presented here demonstrating the performance of ProbCache compared to CEE might seem different than those presented in the original ProbCache paper~\cite{psarasprobabilistic}. The reason behind such difference might be the difference in used network topology, and the size of the cache used. In the first level of our experiments, we have used very small caches to demonstrate that even with small caches it is possible to cache significant amount of content if the space is utilized efficiently. However, small cache sizes are not that favorable to ProbCache or CEE as they only try to cache on path, and the total cache space on path is not big. Thus, only a few of the more popular content items get to stay in the cache, and those also gets swapped out regularly due to the traffic distribution. This results in the below par performance curves for ProbCache and CEE. Please also note, preliminary versions of some of the following results have been presented in an earlier shorter version of the paper~\cite{myinfocom}.

\subsection{Cache Storage Efficiency}

Cache storage efficiency, in the context of this work, refers to the ability of caching unique objects in the network and avoiding duplicates without requiring exhaustive global cooperation. Usually, popular objects are duplicated in the cache network many times causing the cache network to cache far less amount of unique data than its capacity. We argue that the more content the network can hold within it, the better performance it is delivering cache-availability-wise. In this kind of network, the cache network takes responsibility of keeping the popular data alive instead of the content server. While there are other parameters to consider for cache performance (such as latency, network traffic, etc), however, our focus is on the cache retention here, and we will take a look at the other parameters and the effect on those as well.

\begin{figure*}[t]
\begin{minipage}[b]{0.46\textwidth}
\centering
    \includegraphics[width=\textwidth] {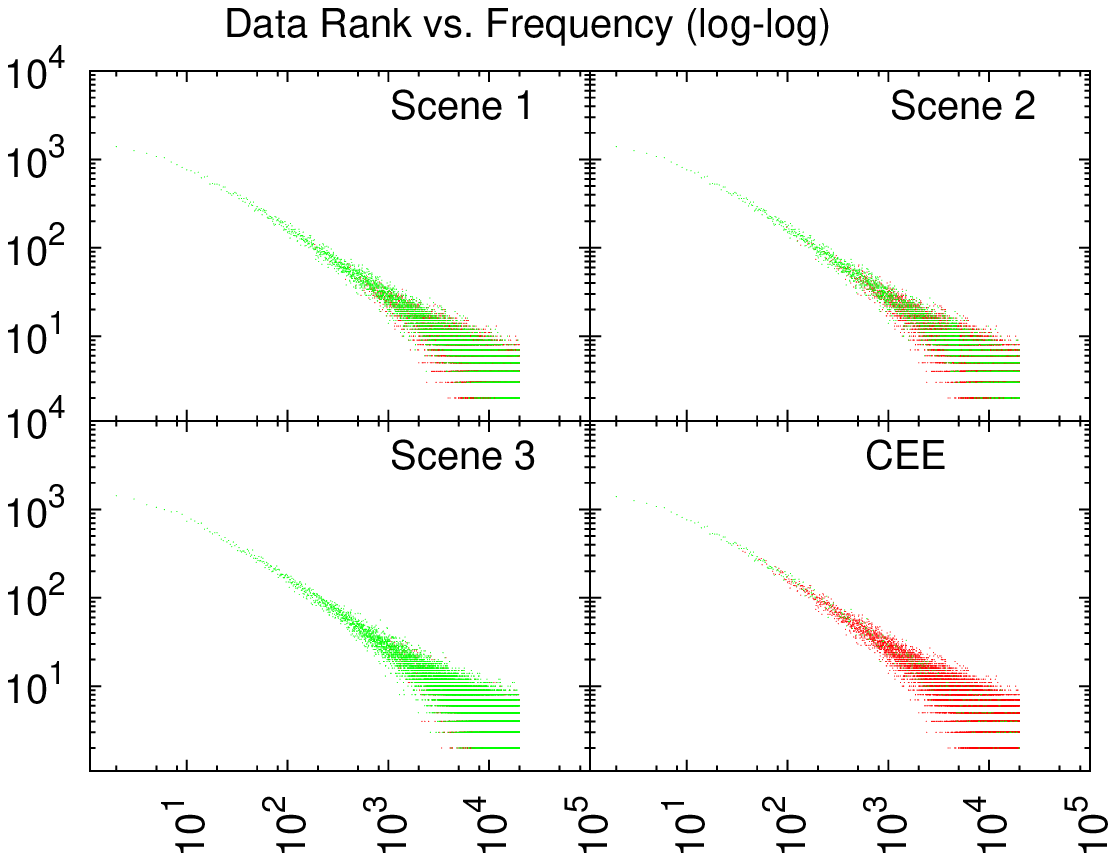}
    \caption{Retention of unique data objects in cache}
    \label{fig:network_cacheretention}
\end{minipage}
\hspace{0.2cm}
\begin{minipage}[b]{0.46\textwidth}
\centering
    \includegraphics[width=\textwidth] {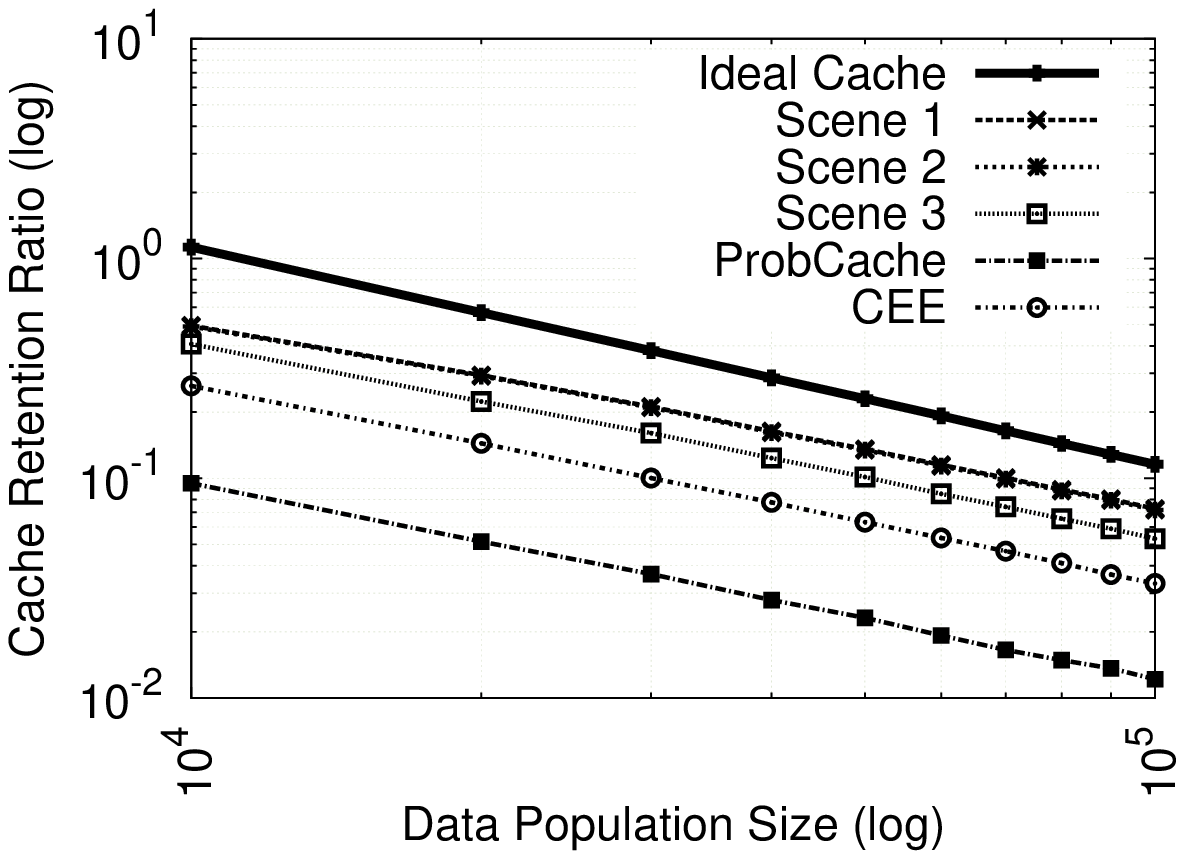}
    \caption{Cache retention ratio with variable data population~\cite{myinfocom}}
    \label{fig:poolsize_cachepercentage}
\end{minipage}
\end{figure*}

To measure this parameter, a network of 20 ASes each containing a 100 routers was built. The topology was created using the Waxman~\cite{waxman1988routing} algorithm. Each router had a cache capacity of 5 data objects (experiments with large numbers are also presented later), resulting in a total of $5\cdot100\cdot20$ = 10,000 cache slots (network capacity $= n_c$) for the whole network and $5\cdot100 = 500$ cache slots for each AS. A data pool of 20,000 data objects (henceforth, data population $= n_p$) was set up. The AS capacity thus is $2.5\%$ of the total population while a router capacity is $0.025\%$ of the population. This cache capacity is quite conservation compared to the setup presented in~\cite{fayazbakhsh2013less}. In their case, derived from the real world traces of various CDN provisioning, the cache budget for each router is $5\%$ of the population. However, the same work also shows that with smaller cache size most of the contemporary caching policies become ineffective, while increasing cache size has non-linear relationship with increasing effectiveness. With a sufficiently large cache per router ($>10\%$), the edge caches start to account for a significant fraction of teh requests and the utility of interior caches becomes marginal. In our experiments, we strive to show the effectiveness of our proposal even with very small cache compared to~\cite{fayazbakhsh2013less}. The data pool is served by several servers across the network. Request for these data are generated 40,000 times according to Zipf-Mandelbrot distribution. Although the number of requests generated might seem low compared to the universe of data objects, the objects in the tail of the heavy tailed distribution are also requested and stored in the cache as shown in Fig.~\ref{fig:network_cacheretention}. This demonstrates the capability of different caching policies to respond to low frequency objects.

\textbf{Network--wide measurements: }After executing the test, the data objects that were found in caches across the network were marked. Fig.~\ref{fig:network_cacheretention} shows the data objects mapped according to their popularity rank in a log-log scale (i.e. rank according to the number of requests on one axis, and the number of cache hits on another), and marks the objects found in the cache with green dots and the ones that are not found with reds. We observe that the proposed algorithms performed consistently better by retaining more objects (more green dots) than CEE in both popular (top left) and unpopular (bottom right) regions. This figure demonstrates that objects ranging from high frequency to low frequency (tail end) were requested during the simulation, and while the on-path caching policies such as CEE can only deal with the high frequency objects (green dots on the top left only), the proposed policies can cache even the low frequency tail end objects (green dots also in the bottom right region). 

\begin{figure*}[t]
\begin{minipage}[b]{0.46\textwidth}
\centering
    \includegraphics[width=\textwidth] {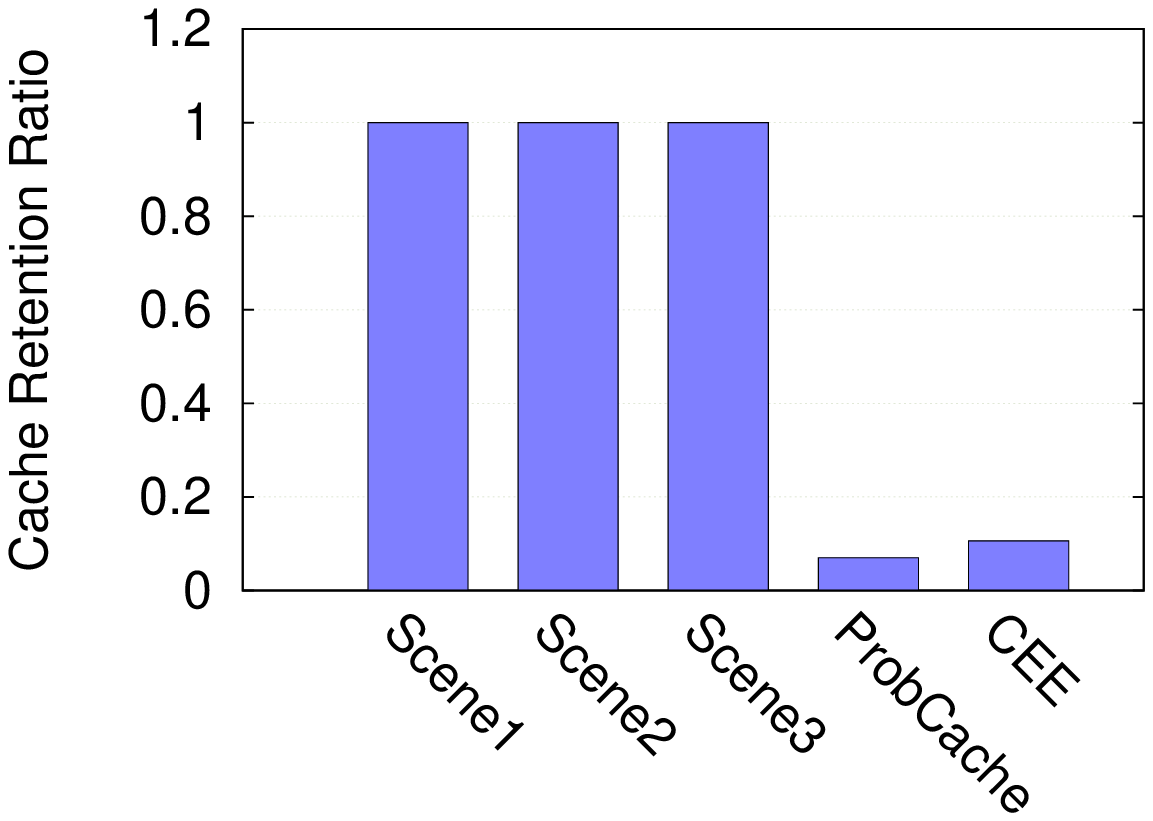}
    \caption{Cache retention: Protocol-wise median}
    \label{fig:as_cacheretention}
\end{minipage}
\hspace{0.2cm}
\begin{minipage}[b]{0.46\textwidth}
\centering
    \includegraphics[width=\textwidth] {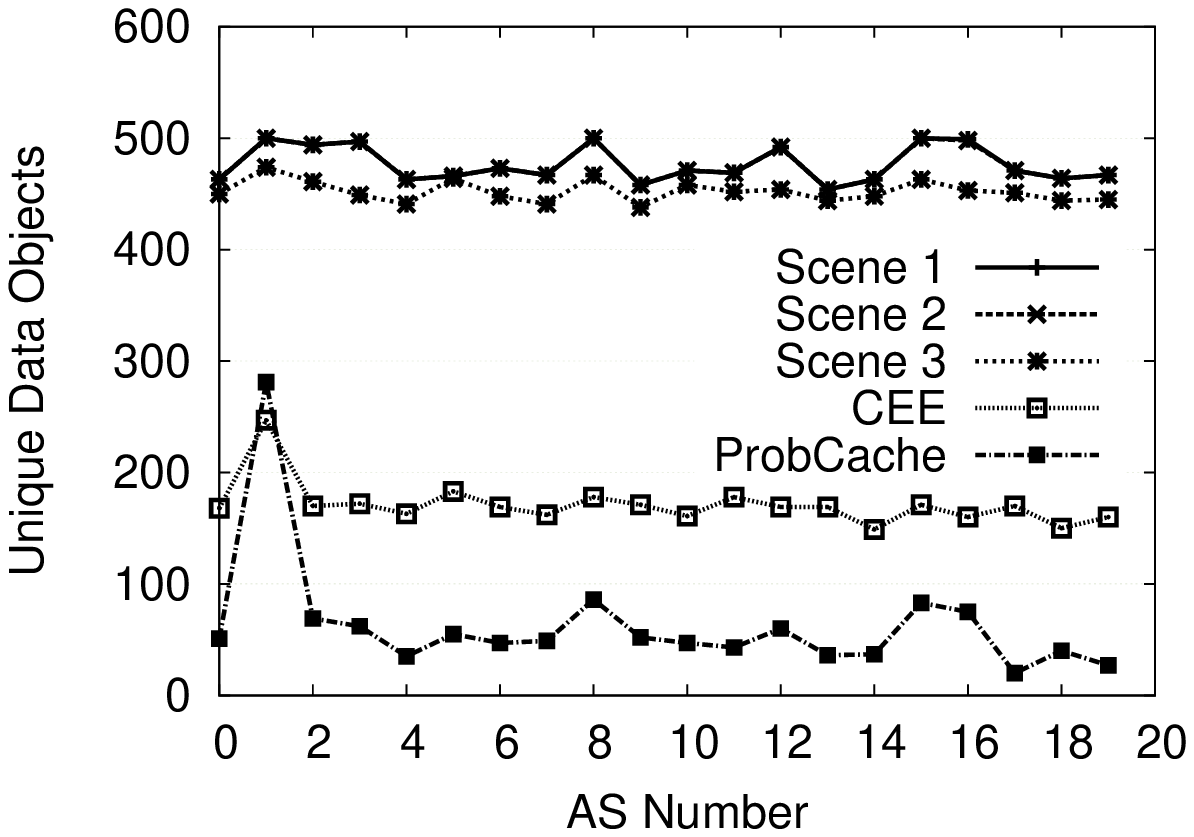}
    \caption{Cache retention: Per AS capability}
    \label{fig:as_uniqdata} 
\end{minipage}
\end{figure*}

To visualize how the network capacity and data population affect this cache storage efficiency parameter, we ran the same experiment with variable $n_p$. We started with a constant $n_c$ and varied $n_p$ within the range $[n_c, (10 \cdot n_c)]$, and each time we generated $2 \cdot n_p$ of data requests. After each run the total number of unique data objects in the network was collected to calculate the cache retention ratio: $\frac{D_q}{D_p}$, where $D_q$ is unique data objects in the cache and $D_p$ is the unique data objects passing through the AS. Fig.~\ref{fig:poolsize_cachepercentage} is a log-log chart of the data population versus the cache retention ratio. The data series marked as \textit{``ideal cache''} signifies the situation where the network is storing only unique values without any duplicates\footnote{Cache retention with ideal cache = $\frac{n_c}{n_p}$}. The figure shows that the proposed scenarios of caching algorithms consistently outperforms CEE and probcache caching. The noticeable information in this figure is that the non-optimal path scenario performed worse than those following the optimal path, although generally we were expecting this scenario to retain more unique data than others due to better cache spreading policy and longer path. This might be because of excess cache thrashing in the longer data path.

\textbf{AS--level measurements: }AS-level cache retention efficiency provides us with the insight of how efficient each AS is on serving the client requests from its own cache. Let's assume a scenario where a certain AS has decided to adopt this efficient cache management policy and does not have any cooperation with any other AS yet. However, the AS knows what content object range its customers are interested in, and it want to cache as much as possible within the AS. We conducted an experiment where $n_p$ was restricted to the same number as a single AS's capacity (500 data objects), which represents the scenario that the AS's intended interest range is the same as its cache space. The AS-wise median cache retention ratio is shown in Fig.~\ref{fig:as_cacheretention}. As can be seen, all three scenarios of the proposed algorithm could achieve fully unique spread of data, while CEE and ProbCache retained significantly lower proportion of the population. This evidently happened due to the heavy duplication of popular data in CEE and in ProbCache. Additionally, we observe the cache retention status of each AS in Fig.~\ref{fig:as_uniqdata}. It is apparent that for a small network and a almost zipf-like data request distribution, optimal path scenarios behave similarly, while non-optimal path behaves a little worse, and others (CEE, Probcache) achieve far less efficiency.

\textbf{Cache fairness: }The observations of the cache spread can be further strengthened by measuring a parameter called the Jain index\cite{jainindex}. This index evaluates how fairly the caches in a network are being used. Experiment shows that the proposed scenarios achieve excellent fairness index in all the ASes, while, both CEE and ProbCache has quite low Jain index. Detailed result is presented in~\cite{myinfocom}. Although Jain index is not usually used for this purpose, it enables us to show the optimal spread of data among the routers of an AS and that the algorithm reduces the possibility of creating a hotspot within the network due to popular routes or placement of the server. 

\begin{figure*}[t]
\begin{minipage}[b]{0.46\textwidth}
\centering
    \includegraphics[width=\textwidth] {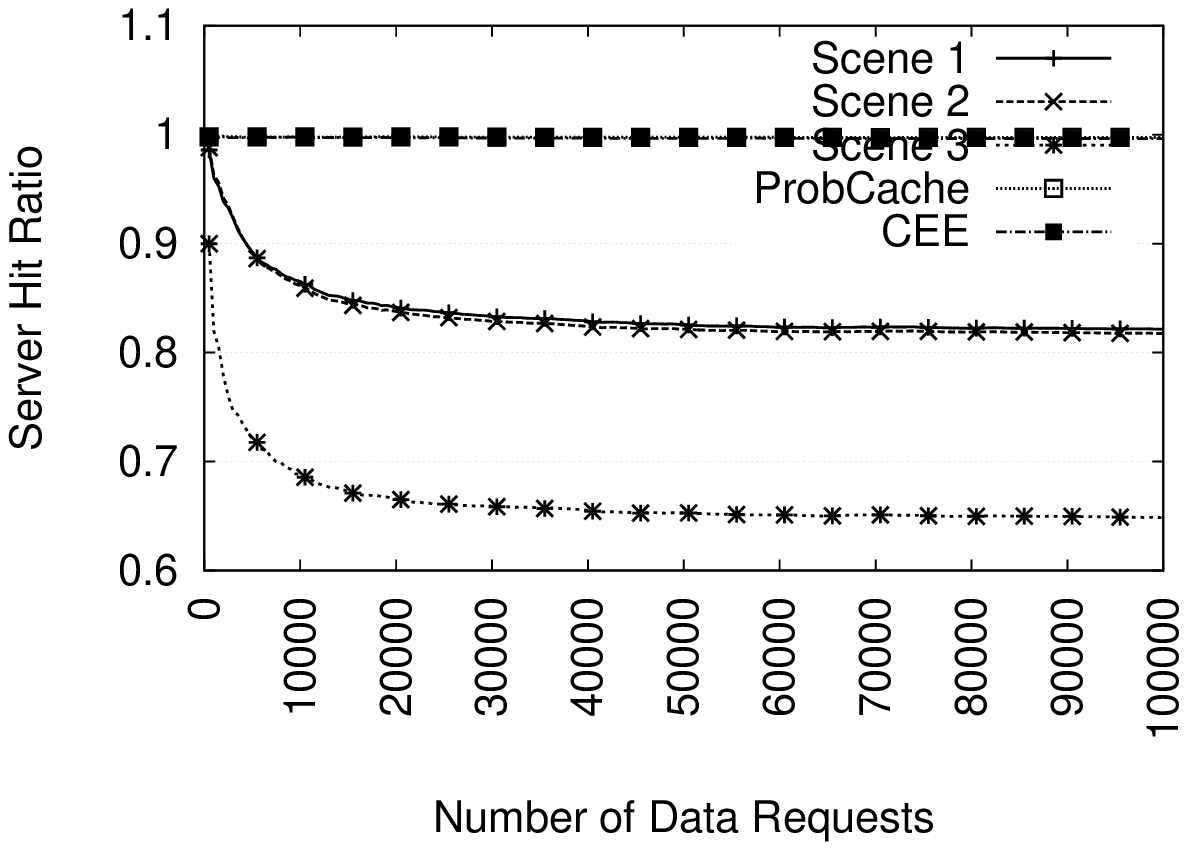}
    \caption{Server hit ratio~\cite{myinfocom}}
    \label{fig:serverhit_ratio}
\end{minipage}
\hspace{0.2cm}
\begin{minipage}[b]{0.46\linewidth}
\centering
  \includegraphics[width=\textwidth] {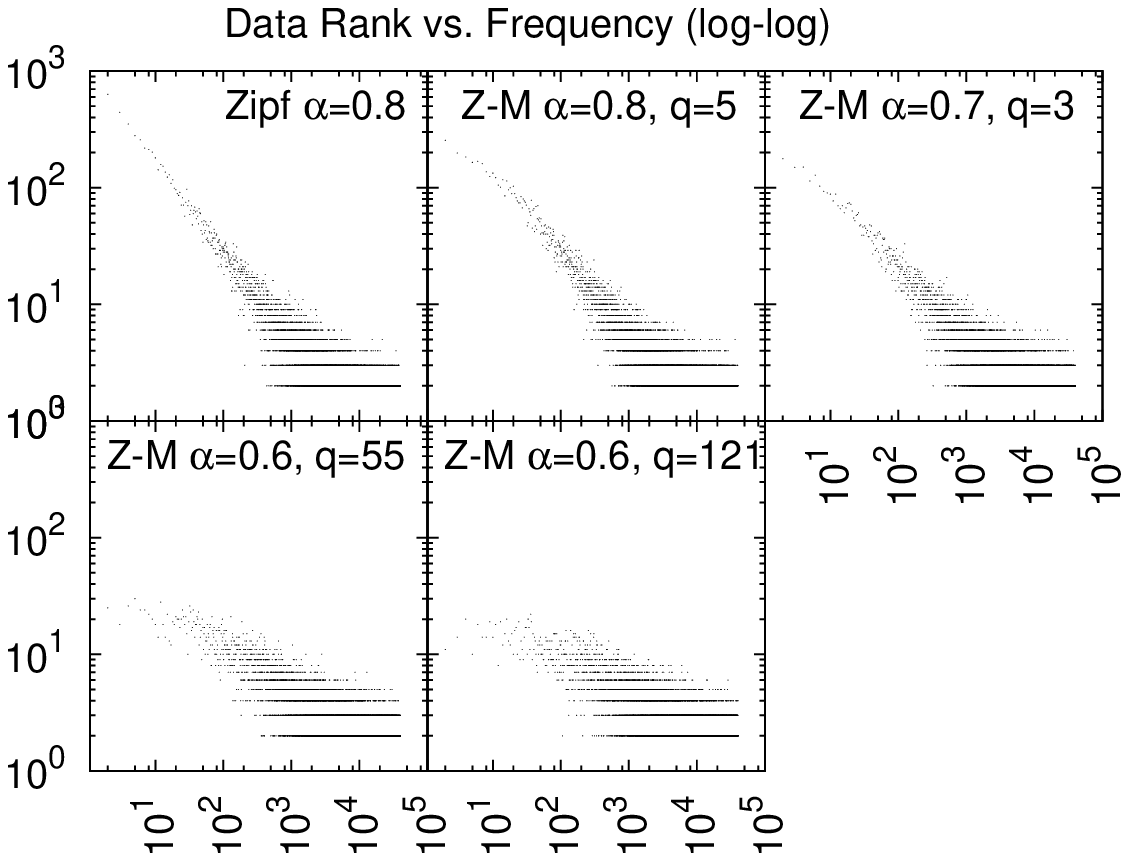} 
    \caption{Log-Log plot of different traffic distributions (datarank vs frequency)} 
    \label{fig:datarank-variation} 
\end{minipage}
\end{figure*}
\begin{figure*}[t]
\begin{minipage}[b]{0.46\linewidth}
\centering
  \includegraphics[width=\textwidth] {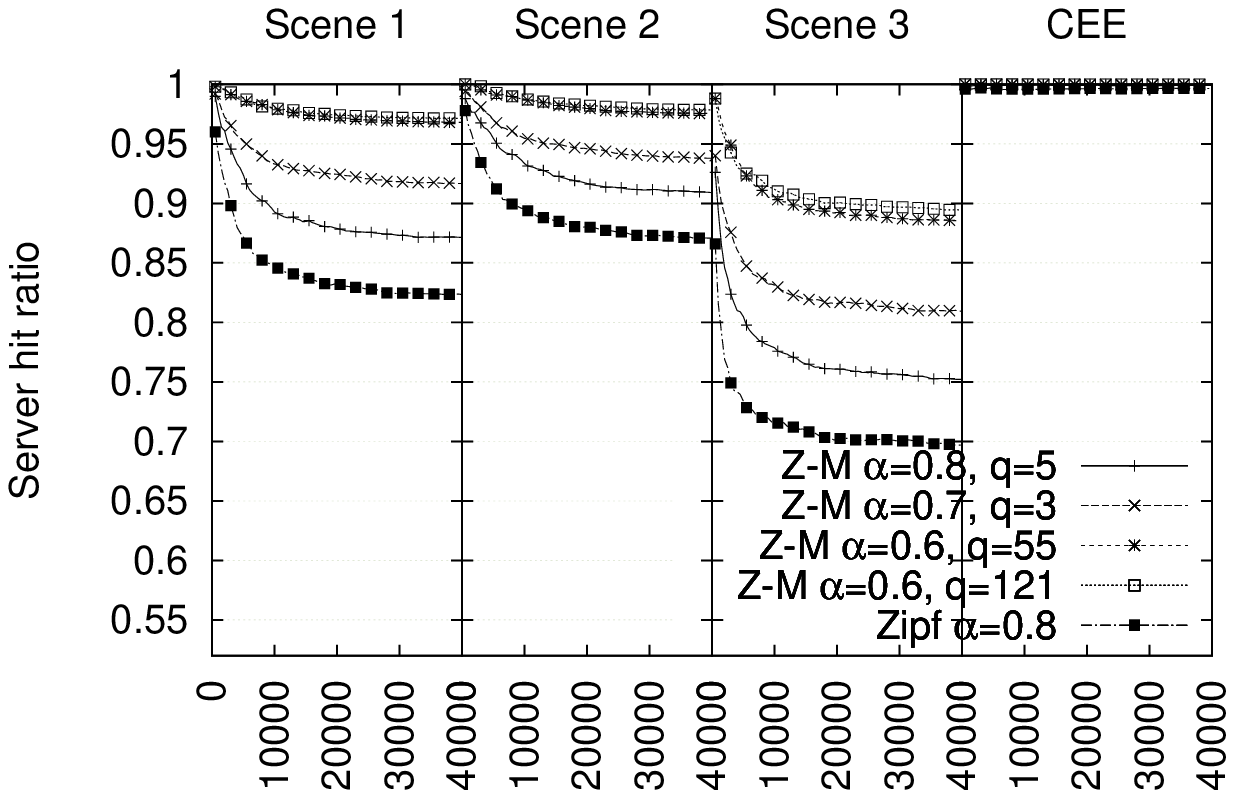} 
    \caption{Server hit ratio for different traffic distributions} 
    \label{fig:zipf-mandel-variation} 
\end{minipage}
\hspace{0.1cm}
\begin{minipage}[b]{0.46\linewidth}
\centering
    \includegraphics[width=\textwidth] {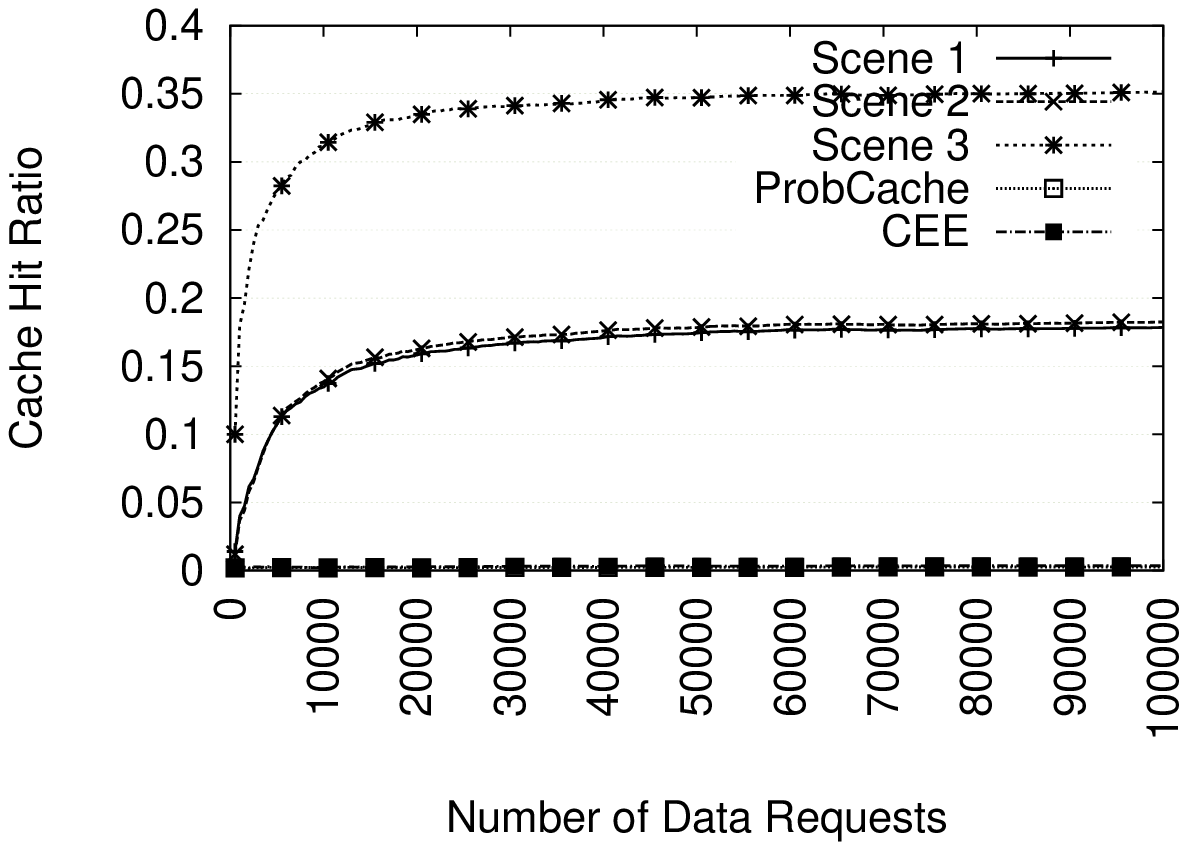}
    \caption{Cache Hit ratio}
    \label{fig:cachehit_ratio}
\end{minipage}
\end{figure*}

\subsection{Server Hit Ratio}
\label{sec:serverhitratio}
The server hit ratio~\cite{chai2012cacheless} defines how much of the data requests the original server has to serve in a caching network in contrast to a network with no caching. The parameter is defined as: $\frac{W_s}{W_t}$, where $W_t$ is the total number of requests generated, and $W_s$ is the number of server hits. For this experiment, the network parameters are as follows, 20 ASes, 100 routers per AS, $n_c=10,000$, and $n_p=20,000$. Fig.~\ref{fig:serverhit_ratio} reveals that with continuous data requests, the proposed algorithms are performing better than both traditional CEE or ProbCache, because the server hit ratio are decreasing at a much faster rate for the proposed algorithms. This result demonstrates how well the proposed schemes can retain and retrieve cached data in a network. With the amount of cache configured for each router in the simulation (which is quite small but still is more than enough considering the total data population), protocols such as CEE and ProbCache cause too much thrashing in the caches. Due to this, most of the requests end up getting served from the server resulting in almost a straight line curve in the plot. 

As it is becoming more and more difficult to predict the Internet traffic distribution (due to novel content sharing paradigms such as P2P, and VoD), we have evaluated our proposed architecture in changing traffic pattern. We have generated traffic according to a zipf distribution with $\alpha = 0.8$, and also Z-M distribution with the combinations $\alpha=0.8, q=5$; $\alpha=0.7, q=3$; $\alpha=0.6, q=55$; and $\alpha=0.6, q=121$; where $q$ is the flatness parameter. All of these values are taken from real world traffic measurements~\cite{hefeeda2008traffic}. Fig.~\ref{fig:datarank-variation} depicts in a log-log scale different traffic patterns using the models used. This figure demonstrates how the difference of popularity between the highly popular and not-so-popular objects are changing with the given parameters. A more elaborate description of such plots can be found here\footnote{http://www.hpl.hp.com/research/idl/papers/ranking/ranking.html}. Fig.~\ref{fig:zipf-mandel-variation} shows the performance of the algorithms with different traffic patterns as in Fig.~\ref{fig:datarank-variation}. Both Phase 1 and 2 performs better with sharp peaked distributions, which is understandable as with flatter distributions popular content amount goes beyond the cache capacity, causing more server hits. On the other hand, ProbCache and CEE performs very similar to a no-cache environment, even in the presence of a small cache. 

\subsection{Cache Hit Ratio}
Cache hit ratio signifies what ratio of the total data is served out of cache instead of reaching the server. Basically, this parameter is the opposite of the server hit ratio parameter that we have presented in Subsection~\ref{sec:serverhitratio}. However, for the sake of completeness we have plotted this parameter in Fig.~\ref{fig:cachehit_ratio}. As expected, the proposed algorithms are performing better and the cache hit ratio increases at a high rate due to a good distribution of data in the available caches, and a proper routing of the data requests to the correct cache. 

\subsection{Cache Latency or Hop Count}
\begin{figure*}[t]
\begin{minipage}[b]{0.46\linewidth}
\centering
    \includegraphics[width=\textwidth] {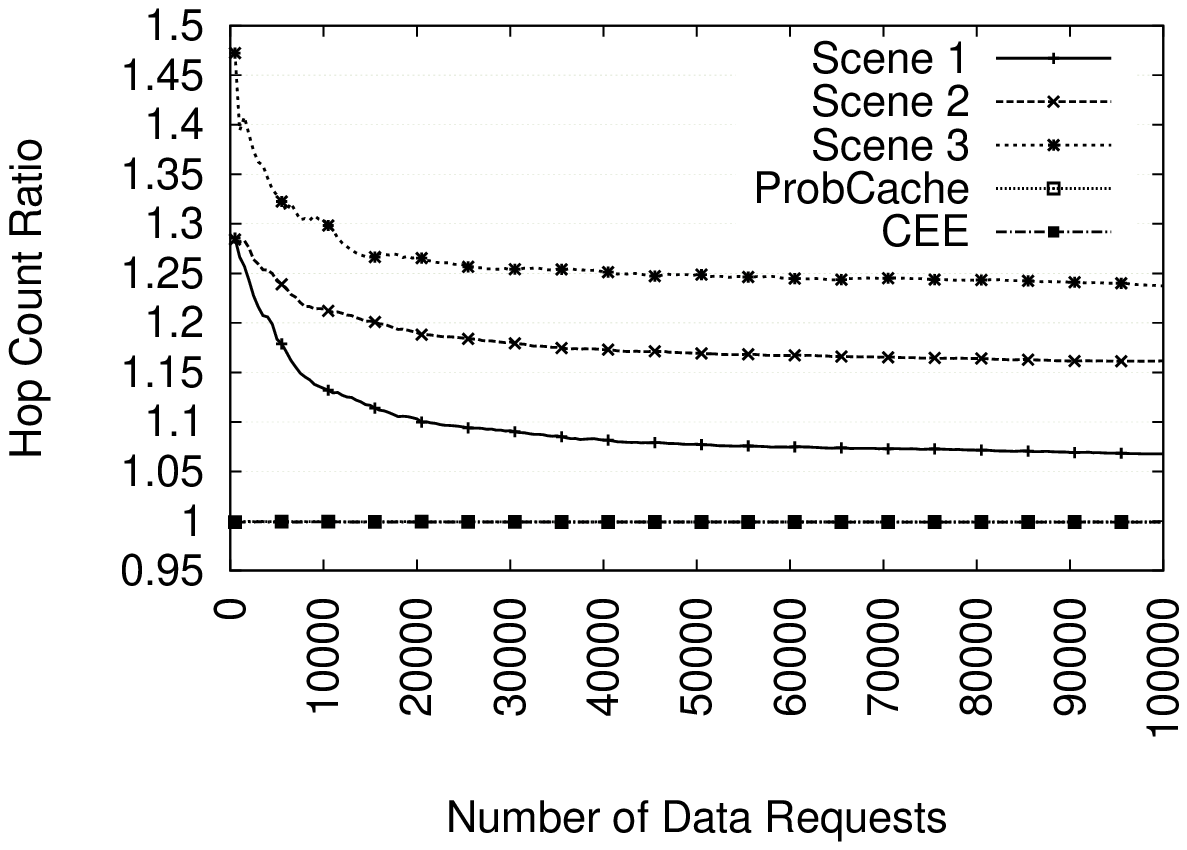}
    \caption{Hopcount ratio}
    \label{fig:hopcount_ratio}
\end{minipage}
\hspace{0.1cm}
\begin{minipage}[b]{0.46\textwidth}
\centering
  \includegraphics[width=\textwidth] {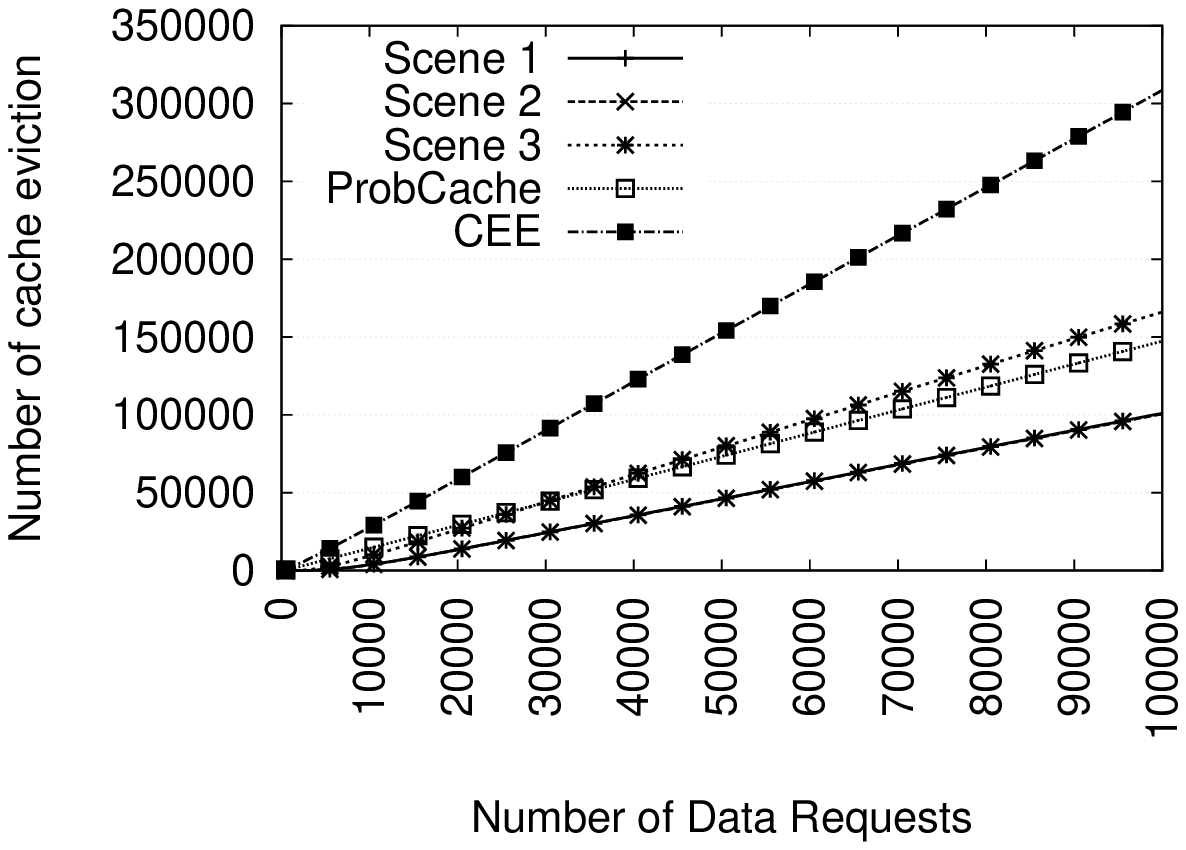}
    \caption{Cache eviction with constant capacity and population~\cite{myinfocom}}
    \label{fig:cache_evict}
\end{minipage}
\end{figure*}

The side effect of the proposed architecture is the possibility of increased hop count due to the detour of the resource requests through non-optimal routes. However, the increase in cache hits due to the increased presence of contents in the cache could prevent much of the extra latency coming out of this detour given there is enough cache space to avoid cache thrashing. Fig.~\ref{fig:hopcount_ratio} shows the hopcount ratio for the same algorithms as in the previous experiments. Hopcount ratio refers to the ratio of required hops for a response to come back to the end user versus the shortest paths hop count from the end user to the server of that particular content. From the figure, we can see that while scenario 3 requires the most average hop counts for data delivery compared to CEE and Probcache, the statistic comes down closer to the shortest path with the warming up of the caches. While it starts with around $50\%$ more required hops compared to the traditional CEE caching, scenario 3 quickly decreases down to only $25\%$ more hops than CEE with quite a small amount of in-node cache. Additionally, scenarios 1 and 2 (which only do in-AS detours) also perform worse compared to CEE and Probcache, and also come down with the warming up of the caches due to more cache hits (to as close as around $5\%$ extra in case of scenario 1). On the other hand, the statistic remains largely the same for Probcache and CEE as the warming up of the cache does not reflect by storing more and more data in the cache, rather it ends up having duplicated data in many places. 

This statistic (in addition to another one presented later for AS level hop count) shows the possible side effects of the proposed algorithm as a trade-off of having a more distributed cache storage mechanism. While the proposed methods utilize the available cache spaces to reduce the dependency on the content server for popular objects and increase the possibility of a cache hit, they also increase the average hop count needed to deliver the object due to detours as a side effect. However, our take from the measurement is that with proper cache configuration it is possible to come quite close to the optimal network latency due to high cache hits. 

\subsection{Network Traffic}
The effect of the proposed method over the overall network traffic pattern is not directly measured, however, it can be inferred from the existing measurement plots such as the hop count (Fig.~\ref{fig:hopcount_ratio}) and the request/response rate. For generalization, if we assume that all the responses are of the same size, then total traffic load for a particular network can be calculated by multiplying the average hop count by the response size and the value should follow the hop count measurement plot. Therefore, inferring from the previous subsection, it can be said that while the increase in the network traffic due to the detours can be significant in the beginning, it quickly goes down as the caches warm up.

\subsection{Cache Eviction}
Natural traffic flow in a network causes churn in the internal caches. Cache eviction is a parameter to measure how much churn a network has due to that. Cache eviction results in more memory accesses and higher processing from the caching entities. Thus, minimizing the cache eviction process while keeping the cache hit rate in par is the best possible outcome for a caching algorithm in ICN for our scenarios. Algorithms that concentrates traffic to certain fixed paths and tries to cache only in certain places (e.g. Scenario 2, Probcache) tend to perform better in this parameter (Fig. \ref{fig:cache_evict}). Algorithms caching everywhere (CEE) and caching in a redirected path (Scene 3) performed worse than others. The reason why Scenario 3 performs a little worse than the others is because of the higher AS hop count, which caused more caches to be queried and replaced along the path.

This parameter also demonstrates the reasons behind high thrashing in the cache system for some algorithms, which eventually cause even the popular objects to get regularly thrown out of the cache. 

\begin{figure*}[t]
\begin{minipage}[b]{0.46\textwidth}
\centering
  \includegraphics[width=\textwidth] {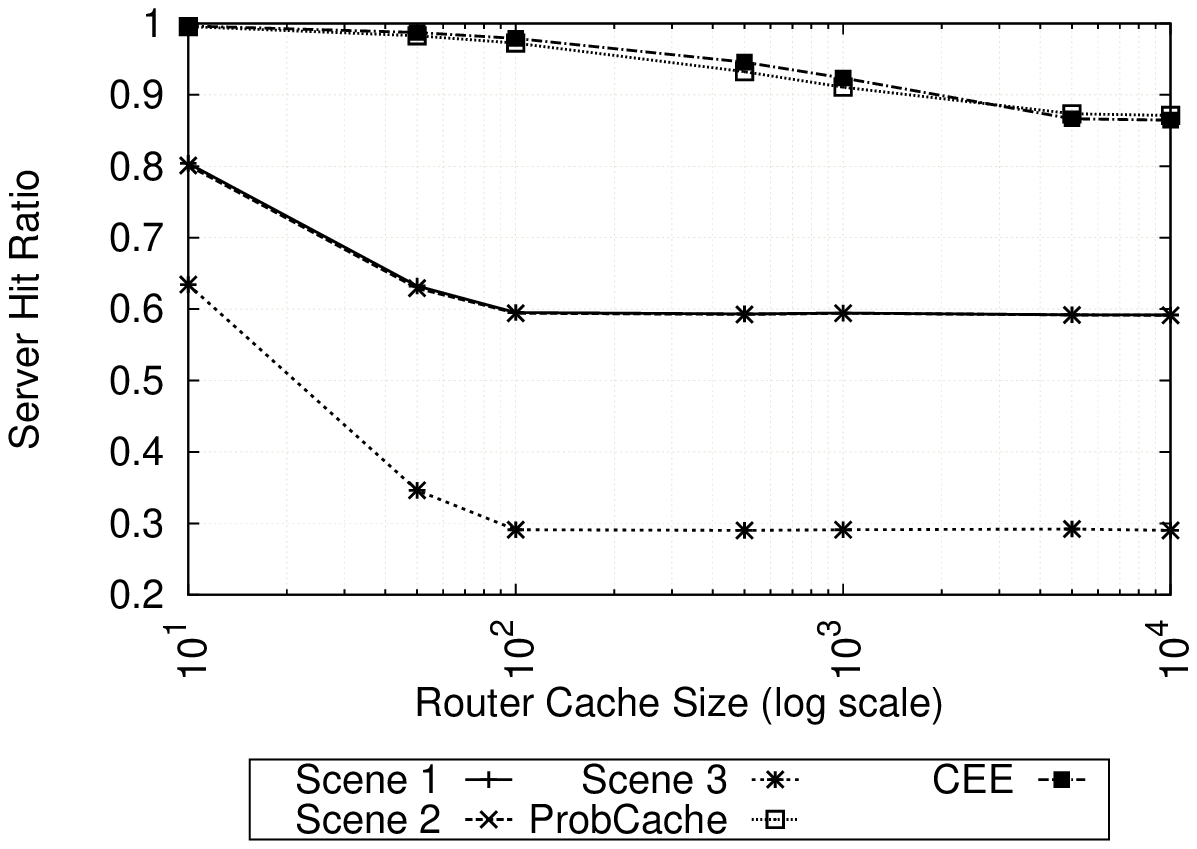}
    \caption{Server hit ratio with variable router cache size~\cite{myinfocom}}
    \label{fig:cachesize_variable}
\end{minipage}
\hspace{0.2cm}
\begin{minipage}[b]{0.46\linewidth}
\centering
    \includegraphics[width=\textwidth] {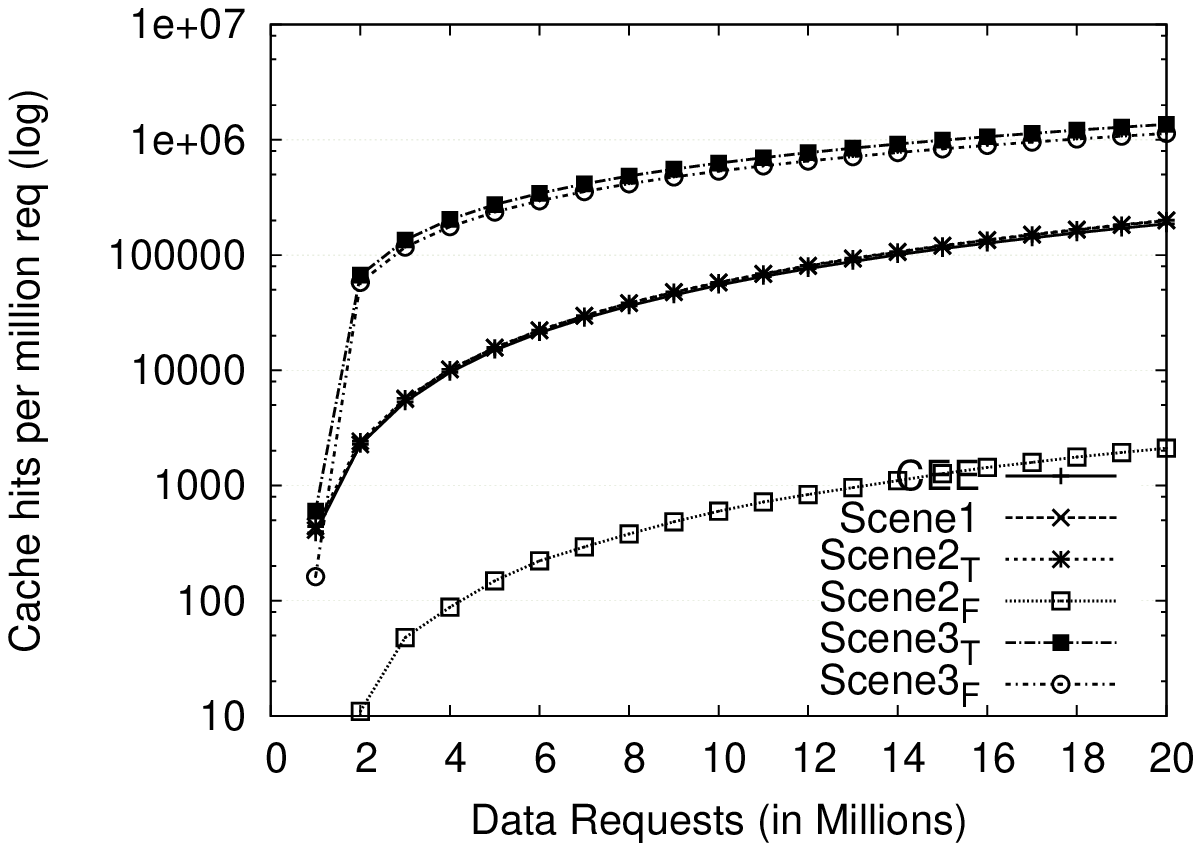}
    \caption{Cache hit per million requests}
    \label{fig:large_cachehit}
\end{minipage}
\end{figure*}
\begin{figure*}[t]
\begin{minipage}[b]{0.46\linewidth}
\centering
    \includegraphics[width=\textwidth] {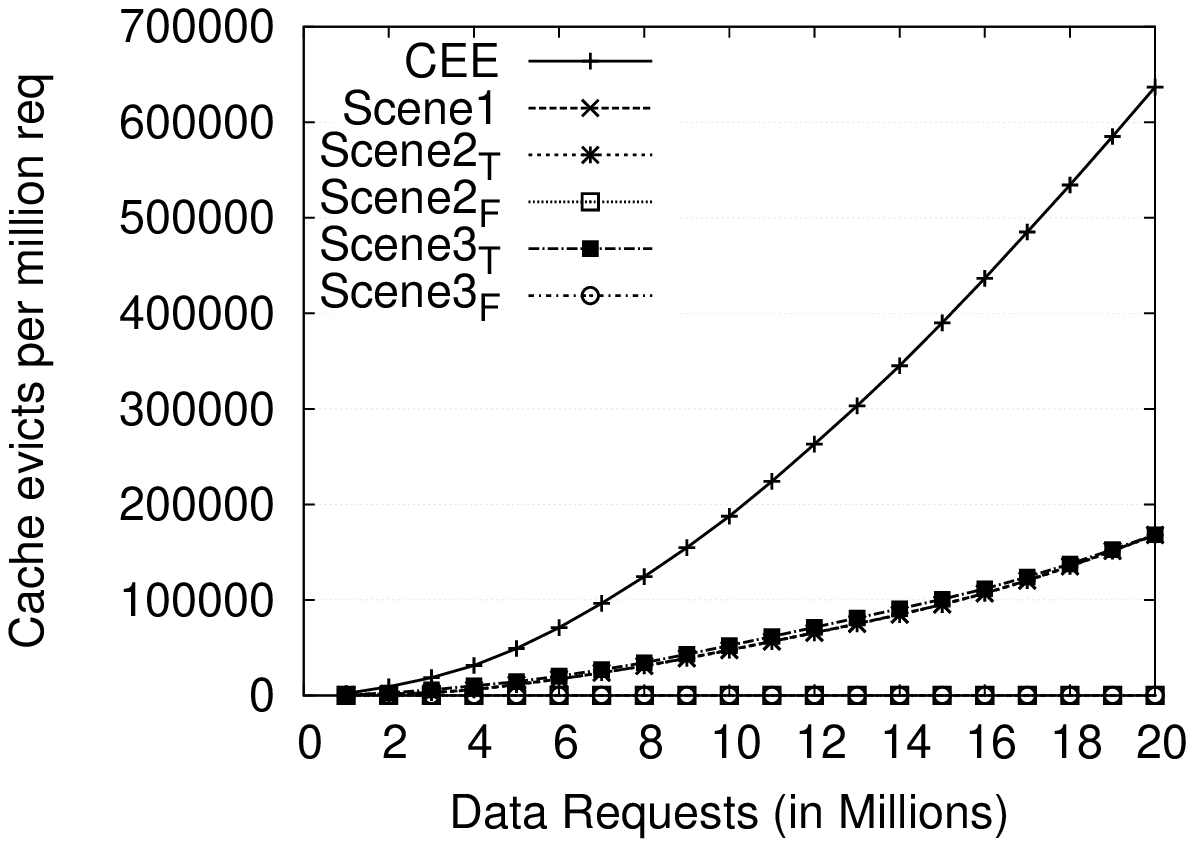}
    \caption{Cache evict per million requests}
    \label{fig:large_evict}
\end{minipage}
\hspace{0.1cm}
\begin{minipage}[b]{0.46\linewidth}
\centering
    \includegraphics[width=\textwidth] {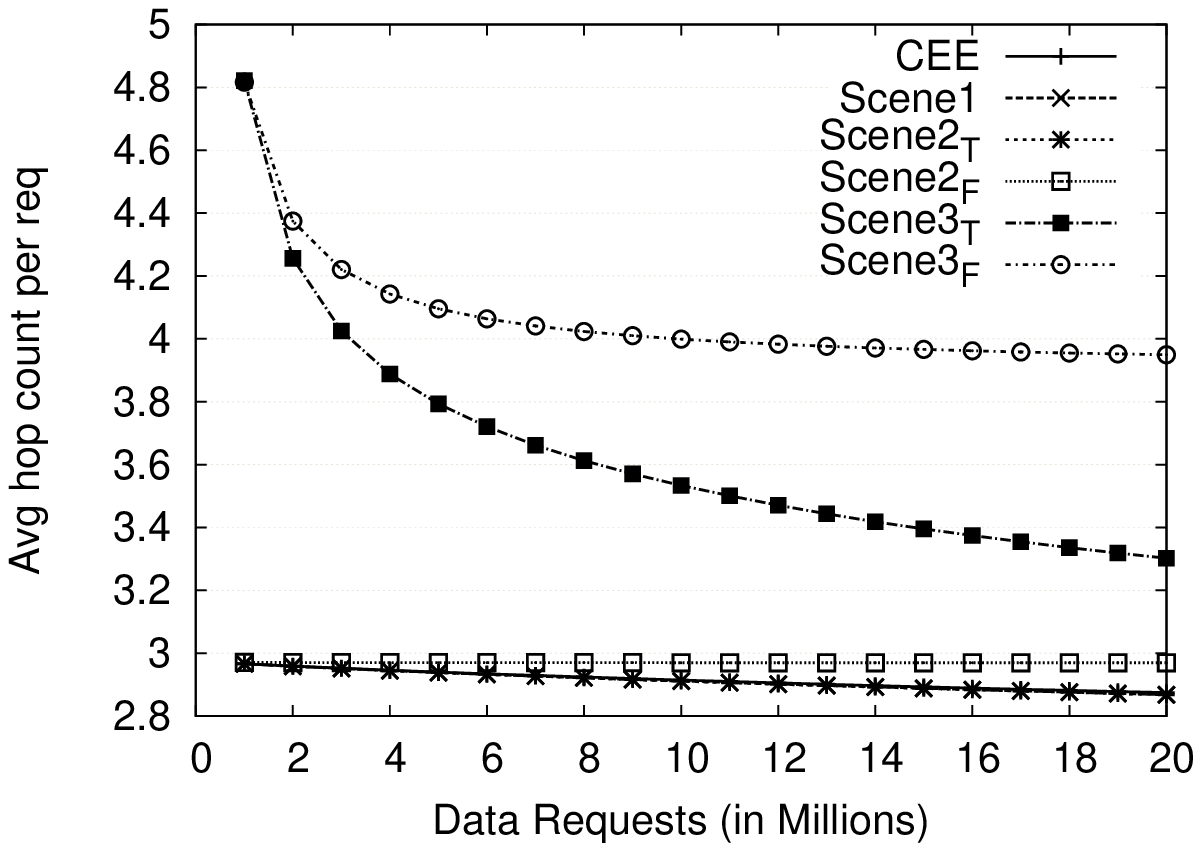}
    \caption{Avg AS-hopcount per request}
    \label{fig:large_hopcount}
\end{minipage}
\end{figure*}

\subsection{Variable Network Parameters}

Although we have experimented with a small cache size to capture the worst case scenario, it would be beneficial to demonstrate the performance of the proposed work under variable cache size. We have varied the router capacity from $10$ to $10,000$ objects\footnote{$n_c = $ Router Capacity $\cdot$ Total routers.} while keeping $n_p$ a constant $40,000$, and ran the experiment for $160,000$ data requests to allow the network to stabilize. After each experiment, we collected the server hit ratio.

From Fig.~\ref{fig:cachesize_variable}, we observe that all proposed scenarios performed better than CEE and ProbCache consistently. Additionally, Scenarios 1 and 2 reach a constant performance already with only 500 cache places per router ($1.25\%$ of the population), while ProbCache and CEE reaches at a constant level with 5000 cache places ($12.5\%$ of the population).

Additionally, to demonstrate our proposal's validity in different kind of topologies, we have generated topologies with a 100 ASes, each containing 516 routers~\cite{Arakawa2012}, using algorithms Waxman, GLP, and BA. The parameters for these algorithms are taken from~\cite{Haddadi2008}, so that the topologies are close to real network. Detailed results are presented in~\cite{myinfocom}, and scenarios 1 and 2 performed consistently better than  both CEE and ProbCache in all three topologies. 

\subsection{Large-Scale Simulation}
To further strengthen our experimental analysis, and additionally, to check whether the proposed algorithm scales up to the real-world Internet, we performed more measurements using the original Internet topology. As we have already laid down a base for the intra-AS performance of the solution, for the measurement, we have only used AS level links from the Skitter measurements of CAIDA~\cite{skitter}. We have treated each AS as one big caching entity where the cache space is as efficient as we found in the previous measurements. For the capacity of each AS, we have again used the Skitter~\cite{skitter} measurements to get the number of routers in each AS, and used that as a base for capacity. As the focus of this section is to demonstrate the scale of the proposed methods, so the comparison with other algorithms were not done extensively. Moreover, as the large scale experiments were done on an AS topology, simulating algorithms such as ProbCache was not possible because it works solely on the router level, which was abstracted away in this phase. For completeness, CEE method is included in the measurement, however, ProbCache is not. The relative performance of ProbCache can be inferred from the experiments above. 

Similar to previous experiments, we generated data requests from a data ID range of 264 Million (based on the total capacity of the network) using Zipf-Mandelbrot distribution from uniformly distributed random source ASes. A total of 20 Million requests were generated from uniformly random sources. We have also increased the scope of our experiment by augmenting the scenarios (as mentioned in Section~\ref{sec:sol}) with different caching decision policies. In both optimal-path scenario 2 and non-optimal-path scenario 3, there are now two different possibilities: (1) only the interested-AS caches and others just pass (SCENE$2_F$ SCENE$3_F$), (2) all the ASes cache (SCENE$2_T$ and SCENE$3_T$). The Figures~\ref{fig:large_cachehit}, \ref{fig:large_hopcount} and~\ref{fig:large_evict} show the performance of the proposed scenarios. 

These large scale simulations reveal very interesting results. As can be seen from Fig.~\ref{fig:large_cachehit}, non-optimal routing resulted in better cache hit than others, while optimal path algorithms (scene 1 and 2) has similar performance except for the one where only the interested AS caches. This shows that even in a very large scale, by distributing the cache, and by routing content requests using the proposed lightweight routing mechanism, it is possible to achieve high cache hit rates. Compared to CEE, the non-optimal scenario performs better in both ways while the optimal path scenario works better only when caching also in non interested ASes. Fig.~\ref{fig:large_hopcount} shows that the cache hit performance of Scene3 comes with the cost of higher AS hops (resulting in higher latency), however, as the figure shows, the difference in hop count gets diminished as the cache hit increases. In the case of SCENE$3_T$, the AS hop difference gets very close to the CEE's hop count (only $0.4$ AS hop difference on average). This reiterates our claim that it is possible to increase the cache retention rate without affecting the hop count much. Fig.~\ref{fig:large_evict} reveals that with selective caching by interested-ASes only, the amount of cache evictions and thrashing is at the minimum. The other proposed methods perform similarly, while CEE performs very poorly in this parameter. 

\section{Discussion}
\label{sec:disc}
In this paper we have described and evaluated a new design architecture for ICN aware caching and routing. While we implemented  and evaluated three scenarios based on the design idea, it primarily should be considered as a basic fabric for efficient caching in ICN. This proposal partially overcomes the problems associated with ICN caching (formulated in earlier sections) by making caching replacement policies flatter. We do it by dividing the whole ID domain into non-competitive sub-domains, while within these sub-domains we still preserve the cache replacement policies, which help to deal with the outliers that are almost never used. The practical design and implementation on top of that fabric can be diverse, of which we have demonstrated three. The results, as shown on the previous sections, are very encouraging. Based on that, a few other possibilities of the architecture are discussed below:

First of all, we required in Scenarios 2 and 3 that each AS claims to be interested for a continuous sector of objects ($[a,b]$). This idea can be extended to imitate the CDN business model, where some ASes advertise data ranges of their business clients and cache for them. In this manner, one of the most successful caching business model of today can be integrated in the ICN network.

Second possibility is one that we consider the most important. As the operators in the current Internet pay to their higher-level providers for the traffic, there are incentives for these providers to create local cache replicators (data center like) and do peering to cooperate. These data centers' unused capacity may be reused within ASes if the price for caching is cheaper than that for forwarding, and an AS can locally decide on forwarding relevant data requests to its own or peer's data center instead of forwarding it upstream through a paid link. While for Tier-1 providers it might not be a lucrative deal to set up such caches, for lower level ISPs (such as Tier-2 and down) the motivation comes from avoiding inter-domain traffic with higher tier and increasing peer traffic which often is free of charge.

Thirdly, in a situation where the links towards the server for a specific content have limited bandwidth, or other constraints such as expensive links, router processing constraints, etc, the proposed solution adds value by spreading and retaining as many data objects as possible in a distributed manner. In the contemporary world, popular objects might be served not by well established companies but by individuals, for whom serving it for an extended period is not often possible. The proposed algorithm addresses this by retaining in the network cache the popular objects even if the source of the data object is not available.

A recent (and parallel to~\cite{myinfocom}) work by Saino et. al.~\cite{saino2013hash} has followed a similar path of research to propose optimal hash-routing schemes to optimize the utilization of available in-network cache space. The work mostly explores the efficacy of different hashing mechanisms to ensure that the cache utilization increases without a significant increase in the network latency or delay. In this regard, we see that the work done in~\cite{saino2013hash} is complementary to our work and can be utilized together to create a complete network-wide solution. 

\section{Conclusion}
\label{sec:concl}
The ICN proposals (e.g. CCNx, DONA, etc), although having many theoretical strengths, suffer from a few fundamental problems such as lack of efficient caching mechanism. Non-cooperative caching, which generally is the proposed caching solution in ICN, is close to the current Internet architecture, and it generally neglects all the ICN benefits. On the other hand, explicit cooperation is hard to achieve and the attempts had negative experiences. In this work, we suggest a novel local cooperation design, which achieves superior performance without generating prohibitive signaling traffic, is lightweight, incrementally deployable, and inherently fault-tolerant.

Our design is presented as an architectural proposal, which can be implemented in many different ways, using various algorithms. As a proof of concept, we implemented three scenarios based on the architectural plan, and evaluated them against the current caching scheme of ICN and a very recent proposal named ProbCache. The results from the evaluation show promising performance in almost all possible parameters, including server hit ratio, cache spread, cache eviction, etc. Reasonable concerns about such indirection based routing is latency and hop count, however, as seen from the large scale simulations, this concern is mitigated by the cache hit increase. Thus, the design has the promise to act as a basis of the ICN caching and routing design that will allow fulfilling the ICN promise of data centrality.

\appendix
\section{Simulator Description}
\label{appn:simulator}
As the primary target of the simulator for this work was to evaluate the performance of the caches in routing elements with different types of routing algorithms, so the functionality of the simulator was limited to finding routes from clients to servers, and simulating a fixed size cache in all the network elements. For this, the simulator reads topologies from text files (nodes and links), creates data structures for nodes and links, and computes routing paths from any node to any other. Request and data packets are routed through the nodes according to the computed paths, and the cache status is updated while passing a node. The cache in a single node uses LRU as a replacement policy. Dijkstra algorithm is used for calculating the shortest path, and a routing decision is made at each hop.

\bibliographystyle{elsarticle-num}
\bibliography{elsevier-icn}

\end{document}